\documentclass[useAMS,usenatbib]{mn2e}

\usepackage{graphicx,color}
\usepackage{amssymb,amsmath}
\usepackage{amsmath} 
\usepackage{multicol}
\usepackage{latexsym}
\usepackage{amsfonts}
\usepackage{amssymb}
\usepackage{wasysym}
\usepackage{textcomp}
\usepackage{longtable,lscape}
\usepackage{mathtools}

\usepackage{aas_macros}
\usepackage[dvipsnames]{xcolor}

\usepackage{natbib}
\usepackage{amssymb,amsmath}

\title[Characterizing low contrast OCs with GAIA]{Characterizing low contrast Galactic open clusters with GAIA DR2}
\author[M. S. Angelo et al.]{M. S. Angelo$^{1}$\thanks{E-mail:
mateusangelo@cefetmg.br}, J. F. C. Santos Jr.$^{2,3}$ and W. J. B. Corradi$^{2,4}$ \\ 
\noindent
$^1$Centro Federal de Educa\c{c}\~ao Tecnol\'ogica de Minas Gerais, Av. Monsenhor Luiz de Gonzaga, 103, 37250-000 Nepomuceno, MG, Brazil\\
$^2$Departamento de F\'isica, ICEx, Universidade Federal de Minas Gerais, Av. Ant\^onio Carlos 6627, 31270-901 Belo Horizonte, MG, Brazil\\
$^3$Departamento de Astronom\'ia, Universidad La Serena, Av. Juan Cisternas 1200, La Serena, Chile\\
$^4$Laborat\'orio Nacional de Astrof\'isica, R. Estados Unidos 154, 37530-000 Itajub\'a, MG, Brazil}

\begin{document}

\date{Accepted XXX. Received XXX; in original form XXX}

\pagerange{\pageref{firstpage}--\pageref{lastpage}} \pubyear{XXXX}

\maketitle

\label{firstpage}

\begin{abstract}


In this study, we characterized 16 objects previously classified as faint or low contrast Galactic open clusters (OCs). We employed parameters associated to the OCs dynamical evolution: core ($r_c$), tidal ($r_t$) and half-mass ($r_{hm}$) radii, age and crossing time ($t_{cr}$). Relations among these parameters were exploited to draw some evolutionary connections. We also included 11 OCs with previous characterizations to provide wider coverage of the parameters space. The investigated sample spans a considerable range in age (log\,($t\,$yr$^{-1}$) $\sim$7.0$-$9.7) and Galactocentric distance ($R_G\,\sim6-11\,$kpc). Most of them present solar metallicity. We employed GAIA DR2 astrometry and photometry and selected member stars through a decontamination algorithm which explores the 3D astrometric space ($\mu_{\alpha}, \mu_{\delta}, \varpi$) to assign membership likelihoods. Previous studies of most of these objects were based mostly on photometric information. All investigated OCs were proved to be real stellar concentrations and relations among their parameters indicate a general disruption scenario in which OCs tend to be more concentrated as they evolve. Internal interactions sucessively drive OCs to develop more dynamically  relaxed structures and make them less subject to mass loss due to tidal effects. Tidal radius tends to increase with $R_G$ in accordance with the strength of the Galactic tidal field. Besides, the correlation between the $r_c$ and the dynamical ratio $\tau_{\textrm{dyn}}=\textrm{age}/t_{cr}$ suggests two distinct evolutionary sequences, which may be consequence of different initial formation conditions.

\end{abstract}


\section{Introduction}


Open clusters (OCs) presenting low stellar density contrast against the Galactic disc field population are particularly challenging to investigate. There are a number of reasons why OCs can be barely distinguishable from the field: $(i)$ they can be projected against a very dense stellar background (e.g., \citeauthor{Ferreira:2019}\,\,\citeyear{Ferreira:2019}), $(ii)$ they can be severely obscured by interstellar absorption (e.g., \citeauthor{Bianchin:2019}\,\,\citeyear{Bianchin:2019}), $(iii)$ they can be intrinsically poorly populated (e.g., \citeauthor{Pavani:2011}\,\,\citeyear{Pavani:2011}; \citeauthor{Angelo:2019a}\,\,2019a) due to the loss of stars during their dynamical evolution.

In this context, the publication of high precision astrometric and photometric data available in the GAIA DR2 catalogue \citep{Gaia-Collaboration:2018} has inaugurated a new era in astronomy. The use of such data allows to circumvent the difficulties arising mainly from points $(i)$ and $(iii)$ above, since the search for statistically significant concentration of stars in the astrometric space can be used to unambigously distinguish cluster from field population. This strategy has led to an increase in the number of OCs discovered and has also allowed more precise characterization of already known ones (e.g., \citeauthor{Cantat-Gaudin:2018a}\,\,2018a; \citeauthor{Castro-Ginard:2018}\,\,\citeyear{Castro-Ginard:2018}; \citeauthor{Monteiro:2019}\,\,\citeyear{Monteiro:2019}; \citeauthor{Ferreira:2019}\,\,\citeyear{Ferreira:2019}). For OCs strongly affected by interstellar extinction, the GAIA visible bands are of limited help, and studies of this kind of object usually require an analysis in conjunction with longer wavelengths.

In the last decade, an effort has been made in the construction of databases compiling lists of OCs and their fundamental parameters. Two of them are WEBDA \citep{Netopil:2012}, containing $\sim1200$ clusters, and DAML02 (\citeauthor{Dias:2002}\,\,\citeyear{Dias:2002}, version 3.5 as of 2016 January), containing $\sim$2200 clusters. \cite{Kharchenko:2013} determined structural, kinematic and astrophysical parameters for $\sim3000$ OCs and associations with the use of an automatic pipeline which works on proper motions and near-infrared photometric data collected from the PPMXL \citep{Roeser:2010} and 2MASS \citep{Skrutskie:2006} catalogues, respectively. More recently, \cite{Bica:2019} presented a catalog of Galactic star clusters, associations and candidates with 10978 entries. For the construction of this multi-band catalog, both large and small catalogues were analysed, together with small sample papers. These catalogues are in general related to specific surveys, such as the ESO-VISTA VVV survey \citep{Minniti:2010}, UKIDSS \citep{Lawrence:2007}, WISE \citep{Cutri:2012} and 2MASS. 

To date, about 3200 objects have measured parameters in the literature and a very limited number of objects have been studied with some detail from a dynamical point of view (e.g., \citeauthor{Piskunov:2007}\,\,\citeyear{Piskunov:2007}; \citeauthor{Piatti:2017a}\,\,\citeyear{Piatti:2017a}; \citeauthor{Angelo:2019b}\,\,2019b). Complete characterizations of OCs including not only fundamental parameters (such as age, distance and interstellar reddening), but also structural and kinematic information, are fundamental for a deeper understanding of the structure, formation and evolution of the Galactic disc \citep{Dias:2019}. Besides, they provide observational constraints for $N-$body simulations aimed at describing the cluster evolution (e.g., \citeauthor{Reina-Campos:2019}\,\,\citeyear{Reina-Campos:2019}; \citeauthor{Rossi:2016}\,\,\citeyear{Rossi:2016}).


The present study is a contribution to increase the number of dynamically investigated objects. Here we focus on a non-exhaustive list of 16 OCs previously classified as low-contrast or faint objects, in general highly field-contaminated and/or poorly-populated. Some of them were indicated as asterisms in previous studies (Section \ref{comments_individual_clusters}). Previous analysis of these objects (\citeauthor{Bica:2005a}\,\,\citeyear{Bica:2005a}; \citeauthor{Bica:2006}\,\,\citeyear{Bica:2006}; \citeauthor{Bica:2011}\,\,\citeyear{Bica:2011}) employed 2MASS photometry and a decontamination tool applied to these clusters' colour-magnitude diagrams (CMDs) in order to determine their physical nature (as real OCs or asterisms). In the present paper, we revisited the analysis of these objects, this time taking full advantage of high-precision GAIA DR2 photometry and astrometry, which allowed more refined lists of member stars. Cleaned CMDs (Section \ref{method}) were obtained after applying a decontamination algorithm which looks for significant overdensities in the astrometric space and assigns membership likelihoods. Since some important discrepancies have been found between our derived parameters and those in the literature, we propose that these OCs' properties should be revised. In Section \ref{comparison_previous_works}, we compared our results for a subsample in common with \citeauthor{Cantat-Gaudin:2018b}\,(2018b).

This paper is organized as follows: in Section~\ref{data_description} we present the investigated sample and the collected data. In Section~\ref{method} we present our analysis method. The results are shown in Section~\ref{results} and discussed in Section~\ref{discussion}, where we characterize the OCs evolutionary stages by employing parameters associated to the OCs dynamical evolution: core ($r_c$), tidal ($r_t$) and half-mass ($r_{hm}$) radii, age and crossing time ($t_{cr}$). Relations between these parameters are exploited to draw some evolutionary connections. Section~\ref{conclusions} summarizes our main conclusions.

\section{Data}
\label{data_description}

Sixteen objects in our sample were taken from the lists of \cite{Bica:2005a}, \cite{Bica:2006} and \cite{Bica:2011}, where they were classified as faint or low-contrast Galactic OCs. In order to enlarge our sample and to provide a wider domain of the investigated dynamical parameters, we complemented our sample with other 11 OCs with previous characterizations in the literature. The list of investigated objects and the whole set of structural and fundamental parameters (as determined according to the procedures stated in Section \ref{method}) are shown in Tables~\ref{struct_params} and \ref{fundamental_params}.

In all cases we employed the Vizier tool\footnote[1]{http://vizier.u-strasbg.fr/viz-bin/VizieR} to extract astrometric and photometric data from the GAIA DR2 catalogue within $\sim1^{\rmn{\circ}}$ of the objects' central coordinates as listed in DAML02. In order to ensure good quality of the astrometric information, we limited the whole set of data to those stars which are consistent with equations 1 and 2 of \cite{Arenou:2018}, aiming at excluding spurious data.

\begin{table*}
  \caption{  Redetermined central coordinates, Galactocentric distances and structural parameters of the studied OCs.  }
  \label{struct_params}
 \begin{tabular}{lcccccccc}

\hline

 Cluster   &$\rmn{RA}$                      &$\rmn{DEC}$                      &$\ell$            &$b$                 & R$_G^{*}$                & $r_c$               & $r_{hm}$                         & $r_t$                      \\
           &($\rmn{h}$:$\rmn{m}$:$\rmn{s}$) & ($\degr$:$\arcmin$:$\arcsec$)   &$(\rmn{^\circ})$  &$(\rmn{^\circ})$    &  (kpc)                   &  (pc)               &     (pc)                         &  (pc)                      \\

\hline
\multicolumn{9}{c}{Main sample} \\
\hline

Haffner\,11                & 07:35:22       & -27:42:03                       & 242.4            & -3.5                &11.1\,$\pm$\,0.7         &3.85\,$\pm$\,0.57    &5.46\,$\pm$\,0.56             &11.54\,$\pm$\,2.14              \\ 
Trumpler\,13               & 10:23:50       & -60:07:41                       & 285.5            & -2.4                & 8.0\,$\pm$\,0.6         &2.30\,$\pm$\,0.36    &3.78\,$\pm$\,0.47             & 9.70\,$\pm$\,1.82              \\
Pismis\,12                 & 09:20:02       & -45:06:54                       & 268.6            &  3.2                & 8.3\,$\pm$\,0.5         &1.15\,$\pm$\,0.14    &2.02\,$\pm$\,0.15             & 5.92\,$\pm$\,1.44              \\
IC\,1434                   & 22:10:29       &  52:51:09                       &  99.9            & -2.7                & 9.0\,$\pm$\,0.5         &2.05\,$\pm$\,0.45    &3.02\,$\pm$\,0.29             & 6.51\,$\pm$\,1.34              \\
Czernik\,32                & 07:50:31       & -29:51:12                       & 245.9            & -1.7                & 9.7\,$\pm$\,0.5         &2.02\,$\pm$\,0.28    &3.47\,$\pm$\,0.36             &10.39\,$\pm$\,1.10              \\
$\textrm{[FSR2007]}$\,1325 & 07:50:33       & -29:57:15                       & 245.9            & -1.7                &10.7\,$\pm$\,0.6         &2.66\,$\pm$\,0.40    &4.32\,$\pm$\,0.43             &11.44\,$\pm$\,1.99              \\
Ruprecht\,130              & 17:47:35       & -30:05:12                       & 359.2            & -1.0                & 6.0\,$\pm$\,0.6         &1.04\,$\pm$\,0.23    &1.43\,$\pm$\,0.23             & 3.08\,$\pm$\,0.41              \\
NGC\,3519                  & 11:04:15       & -61:24:01                       & 290.4            & -1.1                & 7.6\,$\pm$\,0.5         &0.59\,$\pm$\,0.15    &1.03\,$\pm$\,0.16             & 2.96\,$\pm$\,0.99              \\
King\,20                   & 23:33:17       &  58:28:33                       & 112.8            & -2.9                & 8.8\,$\pm$\,0.5         &2.11\,$\pm$\,0.35    &3.21\,$\pm$\,0.33             & 7.30\,$\pm$\,1.26              \\
Haffner\,9                 & 07:24:43       & -17:00:00                       & 231.8            & -0.6                & 9.9\,$\pm$\,0.6         &2.19\,$\pm$\,0.31    &3.60\,$\pm$\,0.25             & 9.07\,$\pm$\,1.54              \\
Czernik\,31                & 07:37:00       & -20:31:05                       & 236.2            &  0.2                & 9.9\,$\pm$\,0.5         &1.76\,$\pm$\,0.29    &2.35\,$\pm$\,0.27             & 4.20\,$\pm$\,0.67              \\
NGC\,2421                  & 07:36:12       & -20:37:48                       & 236.3            &  0.1                & 9.6\,$\pm$\,0.5         &3.43\,$\pm$\,0.50    &4.64\,$\pm$\,0.46             & 8.28\,$\pm$\,1.43              \\
Lynga\,12                  & 16:46:08       & -50:45:04                       & 335.7            & -3.5                & 6.2\,$\pm$\,0.6         &1.03\,$\pm$\,0.24    &1.66\,$\pm$\,0.24             & 4.13\,$\pm$\,1.22              \\ 
Ruprecht\,30               & 07:42:32       & -31:27:20                       & 246.4            & -4.0                &10.1\,$\pm$\,0.6         &1.66\,$\pm$\,0.33    &2.44\,$\pm$\,0.29             & 5.53\,$\pm$\,1.11              \\
Lynga\,2                   & 14:24:14       & -61:21:30                       & 313.8            & -0.4                & 7.4\,$\pm$\,0.5         &      $-$            &      $-$                     & 2.92\,$\pm$\,0.42$^{**}$       \\
Herschel\,1                & 07:47:04       &  00:01:16                       & 219.4            & 12.4                & 8.2\,$\pm$\,0.5         &      $-$            &      $-$                     & 0.86\,$\pm$\,0.09$^{**}$       \\

\hline
\multicolumn{9}{c}{Complementary sample} \\
\hline

NGC\,188                   & 00:47:55       &  85:15:18                       & 122.9             & 22.4               & 9.1\,$\pm$\,0.5         &2.54\,$\pm$\,0.32    &4.71\,$\pm$\,0.38            &15.88\,$\pm$\,2.38              \\
M\,67                      & 08:51:26       &  11:48:00                       & 215.7             & 31.9               & 8.6\,$\pm$\,0.5         &1.63\,$\pm$\,0.23    &3.19\,$\pm$\,0.15            &13.10\,$\pm$\,2.26              \\
NGC\,4337                  & 12:24:05       & -58:07:05                       & 299.3             &  4.6               & 7.2\,$\pm$\,0.5         &1.62\,$\pm$\,0.25    &2.75\,$\pm$\,0.24            & 7.53\,$\pm$\,1.55              \\
NGC\,2439                  & 07:40:45       & -31:41:18                       & 246.4             & -4.5               & 9.9\,$\pm$\,0.6         &2.62\,$\pm$\,0.40    &4.39\,$\pm$\,0.39            &11.90\,$\pm$\,1.51              \\
Collinder\,110             & 06:38:48       &  02:03:49                       & 209.6             & -1.9               & 9.8\,$\pm$\,0.5         &6.34\,$\pm$\,1.20    &7.08\,$\pm$\,0.78            & 9.57\,$\pm$\,1.20              \\
Collinder\,261             & 12:38:06       & -68:22:31                       & 301.7             & -5.5               & 7.0\,$\pm$\,0.5         &3.13\,$\pm$\,0.52    &5.28\,$\pm$\,0.48            &16.10\,$\pm$\,3.06              \\
Czernik\,37                & 17:53:17       & -27:22:10                       &   2.2             & -0.6               & 6.6\,$\pm$\,0.6         &0.57\,$\pm$\,0.08    &1.20\,$\pm$\,0.11            & 5.34\,$\pm$\,1.64              \\
Collinder\,351             & 17:49:02       & -28:44:27                       &   0.6             & -0.5               & 6.3\,$\pm$\,0.6         &0.59\,$\pm$\,0.15    &0.96\,$\pm$\,0.13            & 2.62\,$\pm$\,0.74              \\
NGC\,3680                  & 11:25:39       & -43:13:26                       & 286.8             & 16.9               & 7.8\,$\pm$\,0.5         &0.45\,$\pm$\,0.10    &0.73\,$\pm$\,0.10            & 2.10\,$\pm$\,0.66              \\
NGC\,6216                  & 16:49:24       & -44:43:17                       & 340.7             &  0.0               & 6.1\,$\pm$\,0.7         &1.82\,$\pm$\,0.27    &2.73\,$\pm$\,0.24            & 6.08\,$\pm$\,0.73              \\
BH\,200                    & 16:49:55       & -44:11:15                       & 341.1             &  0.2               & 6.2\,$\pm$\,0.5         &1.08\,$\pm$\,0.17    &1.79\,$\pm$\,0.22            & 4.82\,$\pm$\,0.85              \\

\hline                                                                                                                                                                                                                     
\multicolumn{9}{l}{ \textit{Note 1}: To convert arcmin into pc we used the expression $r(\textrm{pc})=r(\textrm{arcmin})\times(\pi\,/10800)\times10^{[(m-M)_{0}+5]/5}$,}\\    
                          
\multicolumn{9}{l}{ where $(m-M)_{0}$ is the OC distance modulus (see Table \ref{fundamental_params}). } \\

\multicolumn{9}{l}{ \textit{Note 2}: Fourteen of the 16 OCs in the main sample were previously investigated in \cite{Bica:2005a},   } \\

\multicolumn{9}{l}{ \cite{Bica:2006} and \cite{Bica:2011}. The OCs NGC\,2421 and [FSR2007]\,1325 were included since } \\

\multicolumn{9}{l}{   they are projected in the same regions of Czernik\,31 and Czernik\,32, respectively (see Section \ref{comments_individual_clusters}). } \\

\multicolumn{9}{l}{ $^{*}$ The $R_G$ values were obtained from the distances in Table \ref{fundamental_params}, assuming that the Sun is located at 8.0\,$\pm$\,0.5\,kpc }\\

\multicolumn{9}{l}{  from the Galactic centre \citep{Reid:1993a}. } \\

\multicolumn{9}{l}{$^{**}$ No profile fits could be performed. The $r_t$ was assumed as being equal to the cluster's limiting radius }  \\

\multicolumn{9}{l}{ (Section \ref{structural_parameters}). }


\end{tabular}
\end{table*}

\begin{table*}
  \caption{ Fundamental parameters, mean proper motion components and crossing times for the studied OCs. }
  \label{fundamental_params}
 \begin{tabular}{lcccccccc}

\hline

Cluster   &$(m-M)_0$   &$d$   &$E(B-V)$   &log ($t\,$yr$^{-1}$)   &$[Fe/H]$   &$\langle \mu_{\alpha} \rangle$     &$\langle \mu_{\delta} \rangle$    &$t_{cr}$   \\
 &(mag)   &(kpc)   &(mag)         &         &(dex)       &(mas.yr$^{-1}$)    &(mas.yr$^{-1}$)   &(Myr)      \\

\hline                                                                                                                                                                                          
\multicolumn{9}{c}{Main sample} \\                                                                                                                                                              
\hline

Haffner\,11                      & 13.45\,$\pm$\,0.40      & 4.90\,$\pm$\,0.90  & 0.59\,$\pm$\,0.10   & 8.90\,$\pm$\,0.10      & 0.00\,$\pm$\,0.17    & -1.492\,$\pm$\,0.183                          & 3.170\,$\pm$\,0.179                             &1.07\,$\pm$\,0.12   \\ 
Trumpler\,13                     & 13.10\,$\pm$\,0.40      & 4.17\,$\pm$\,0.77  & 0.63\,$\pm$\,0.10   & 8.10\,$\pm$\,0.10      &-0.06\,$\pm$\,0.13    & -6.742\,$\pm$\,0.028                          & 4.104\,$\pm$\,0.090                             &3.32\,$\pm$\,1.26   \\
Pismis\,12                       & 11.48\,$\pm$\,0.40      & 1.98\,$\pm$\,0.36  & 0.55\,$\pm$\,0.05   & 9.20\,$\pm$\,0.10      & 0.00\,$\pm$\,0.17    & -6.677\,$\pm$\,0.211                          & 4.802\,$\pm$\,0.166                             &0.93\,$\pm$\,0.11   \\
IC\,1434                         & 12.43\,$\pm$\,0.30      & 3.07\,$\pm$\,0.42  & 0.49\,$\pm$\,0.10   & 8.50\,$\pm$\,0.10      &-0.06\,$\pm$\,0.20    & -3.895\,$\pm$\,0.035                          &-3.346\,$\pm$\,0.157                             &1.07\,$\pm$\,0.11   \\
Czernik\,32                      & 12.50\,$\pm$\,0.15      & 3.16\,$\pm$\,0.22  & 0.85\,$\pm$\,0.05   & 9.10\,$\pm$\,0.10      &-0.13\,$\pm$\,0.23    & -2.948\,$\pm$\,0.141                          & 2.488\,$\pm$\,0.123                             &1.35\,$\pm$\,0.15   \\
$\textrm{[FSR2007]}$\,1325       & 13.30\,$\pm$\,0.40      & 4.57\,$\pm$\,0.84  & 1.14\,$\pm$\,0.10   & 7.00\,$\pm$\,0.20      & 0.22\,$\pm$\,0.17    & -2.715\,$\pm$\,0.071                          & 3.436\,$\pm$\,0.082                             &1.97\,$\pm$\,0.41   \\
Ruprecht\,130                    & 11.50\,$\pm$\,0.40      & 2.00\,$\pm$\,0.37  & 1.22\,$\pm$\,0.10   & 8.40\,$\pm$\,0.10      & 0.14\,$\pm$\,0.20    &  0.406\,$\pm$\,0.223                          &-1.843\,$\pm$\,0.056                             &0.57\,$\pm$\,0.10   \\
NGC\,3519                        & 11.19\,$\pm$\,0.40      & 1.70\,$\pm$\,0.31  & 0.17\,$\pm$\,0.10   & 8.70\,$\pm$\,0.10      & 0.18\,$\pm$\,0.15    & -6.438\,$\pm$\,0.079                          & 3.116\,$\pm$\,0.069                             &1.24\,$\pm$\,0.27   \\
King\,20                         & 12.11\,$\pm$\,0.40      & 1.73\,$\pm$\,0.32  & 0.89\,$\pm$\,0.10   & 8.20\,$\pm$\,0.20      & 0.28\,$\pm$\,0.18    & -2.667\,$\pm$\,0.128                          &-2.630\,$\pm$\,0.164                             &2.06\,$\pm$\,0.24   \\
Haffner\,9                       & 12.30\,$\pm$\,0.30      & 2.64\,$\pm$\,0.49  & 0.80\,$\pm$\,0.10   & 8.55\,$\pm$\,0.15      &-0.13\,$\pm$\,0.23    & -0.940\,$\pm$\,0.231                          & 0.662\,$\pm$\,0.441                             &0.87\,$\pm$\,0.09   \\
Czernik\,31                      & 11.95\,$\pm$\,0.30      & 2.88\,$\pm$\,0.40  & 0.60\,$\pm$\,0.10   & 8.00\,$\pm$\,0.40      & 0.05\,$\pm$\,0.20    & -1.946\,$\pm$\,0.138                          & 3.035\,$\pm$\,0.130                             &0.98\,$\pm$\,0.14   \\
NGC\,2421                        & 11.60\,$\pm$\,0.40      & 2.46\,$\pm$\,0.34  & 0.55\,$\pm$\,0.10   & 7.75\,$\pm$\,0.30      &-0.13\,$\pm$\,0.23    & -3.148\,$\pm$\,0.108                          & 3.085\,$\pm$\,0.112                             &2.75\,$\pm$\,0.29   \\
Lynga\,12                        & 11.15\,$\pm$\,0.40      & 2.09\,$\pm$\,0.39  & 0.88\,$\pm$\,0.15   & 8.70\,$\pm$\,0.15      & 0.00\,$\pm$\,0.29    & -2.318\,$\pm$\,1.004                          &-3.749\,$\pm$\,0.851                             &0.12\,$\pm$\,0.02   \\ 
Ruprecht\,30                     & 12.90\,$\pm$\,0.40      & 3.80\,$\pm$\,0.70  & 0.63\,$\pm$\,0.10   & 7.50\,$\pm$\,0.20      & 0.00\,$\pm$\,0.29    & -2.026\,$\pm$\,0.193                          & 3.091\,$\pm$\,0.264                             &0.62\,$\pm$\,0.18   \\
Lynga\,2                         &  9.90\,$\pm$\,0.30      & 0.95\,$\pm$\,0.13  & 0.40\,$\pm$\,0.10   & 7.90\,$\pm$\,0.30      &-0.13\,$\pm$\,0.23    & -6.683\,$\pm$\,0.093                          &-4.681\,$\pm$\,0.096                             &      $ - $         \\
Herschel\,1                      &  7.35\,$\pm$\,0.40      & 0.29\,$\pm$\,0.05  & 0.06\,$\pm$\,0.15   & 8.65\,$\pm$\,0.65      & 0.00\,$\pm$\,0.23    &  0.568\,$\pm$\,0.269                          &-4.148\,$\pm$\,0.189                             &      $ - $         \\

\hline                                                                                                                                                                                                                  
\multicolumn{9}{c}{Complementary sample} \\                                                                                                                                                                             
\hline

NGC\,188                         & 11.30\,$\pm$\,0.20      & 1.82\,$\pm$\,0.17  & 0.13\,$\pm$\,0.05   & 9.70\,$\pm$\,0.15     & 0.00\,$\pm$\,0.17     & -2.312\,$\pm$\,0.169                          &-0.940\,$\pm$\,0.081                             &2.72\,$\pm$\,0.23    \\
M\,67                            &  9.45\,$\pm$\,0.20      & 0.78\,$\pm$\,0.07  & 0.06\,$\pm$\,0.05   & 9.60\,$\pm$\,0.10     &-0.06\,$\pm$\,0.13     &-10.982\,$\pm$\,0.154                          &-2.946\,$\pm$\,0.187                             &4.78\,$\pm$\,0.42    \\
NGC\,4337                        & 11.65\,$\pm$\,0.20      & 2.14\,$\pm$\,0.20  & 0.49\,$\pm$\,0.05   & 9.20\,$\pm$\,0.10     &-0.32\,$\pm$\,0.12     & -8.812\,$\pm$\,0.120                          & 1.475\,$\pm$\,0.096                             &1.83\,$\pm$\,0.17    \\
NGC\,2439                        & 12.70\,$\pm$\,0.40      & 3.47\,$\pm$\,0.64  & 0.55\,$\pm$\,0.10   & 7.25\,$\pm$\,0.20     &-0.22\,$\pm$\,0.28     & -2.275\,$\pm$\,0.126                          & 3.146\,$\pm$\,0.137                             &1.53\,$\pm$\,0.14    \\
Collinder\,110                   & 11.57\,$\pm$\,0.20      & 2.06\,$\pm$\,0.19  & 0.50\,$\pm$\,0.05   & 9.25\,$\pm$\,0.10     &-0.13\,$\pm$\,0.23     & -1.093\,$\pm$\,0.145                          &-2.062\,$\pm$\,0.155                             &3.87\,$\pm$\,0.44    \\
Collinder\,261                   & 12.04\,$\pm$\,0.30      & 2.56\,$\pm$\,0.35  & 0.33\,$\pm$\,0.05   & 9.70\,$\pm$\,0.15     & 0.00\,$\pm$\,0.17     & -6.350\,$\pm$\,0.156                          &-2.685\,$\pm$\,0.152                             &2.24\,$\pm$\,0.21    \\
Czernik\,37                      & 10.75\,$\pm$\,0.50      & 1.41\,$\pm$\,0.33  & 1.50\,$\pm$\,0.10   & 8.70\,$\pm$\,0.15     & 0.00\,$\pm$\,0.29     &  0.434\,$\pm$\,0.203                          &-0.447\,$\pm$\,0.143                             &0.81\,$\pm$\,0.09    \\
Collinder\,351                   & 11.15\,$\pm$\,0.40      & 1.70\,$\pm$\,0.31  & 0.87\,$\pm$\,0.15   & 8.85\,$\pm$\,0.10     &-0.06\,$\pm$\,0.33     & -0.237\,$\pm$\,0.162                          &-1.700\,$\pm$\,0.559                             &0.18\,$\pm$\,0.03    \\
NGC\,3680                        &  9.78\,$\pm$\,0.40      & 0.90\,$\pm$\,0.17  & 0.15\,$\pm$\,0.05   & 9.20\,$\pm$\,0.10     &-0.06\,$\pm$\,0.20     & -7.254\,$\pm$\,0.196                          & 1.148\,$\pm$\,0.231                             &0.69\,$\pm$\,0.12    \\
NGC\,6216                        & 11.60\,$\pm$\,0.50      & 2.09\,$\pm$\,0.48  & 1.04\,$\pm$\,0.10   & 7.85\,$\pm$\,0.30     & 0.00\,$\pm$\,0.23     & -1.261\,$\pm$\,0.209                          &-2.535\,$\pm$\,0.135                             &1.48\,$\pm$\,0.14    \\
BH\,200                          & 11.45\,$\pm$\,0.20      & 1.95\,$\pm$\,0.18  & 1.00\,$\pm$\,0.10   & 7.90\,$\pm$\,0.30     & 0.00\,$\pm$\,0.23     & -0.070\,$\pm$\,0.176                          &-1.182\,$\pm$\,0.127                             &1.19\,$\pm$\,0.17    \\

\hline

\end{tabular}
\end{table*}

\section{Method}
\label{method}

\subsection{Pre-analysis: Finding the signature of a cluster}
\label{preanalysis}

In the first step of our analysis, we searched for signals of the physical existence of the investigated clusters by inspecting the dispersion of data in their CMDs and VPDs, previously to the application of the decontamination method described in Section~\ref{decontam_method}. 

\begin{figure*}
\begin{center}

\parbox[c]{1.0\textwidth}
  {
   
   \begin{center}
      \includegraphics[width=0.45\textwidth]{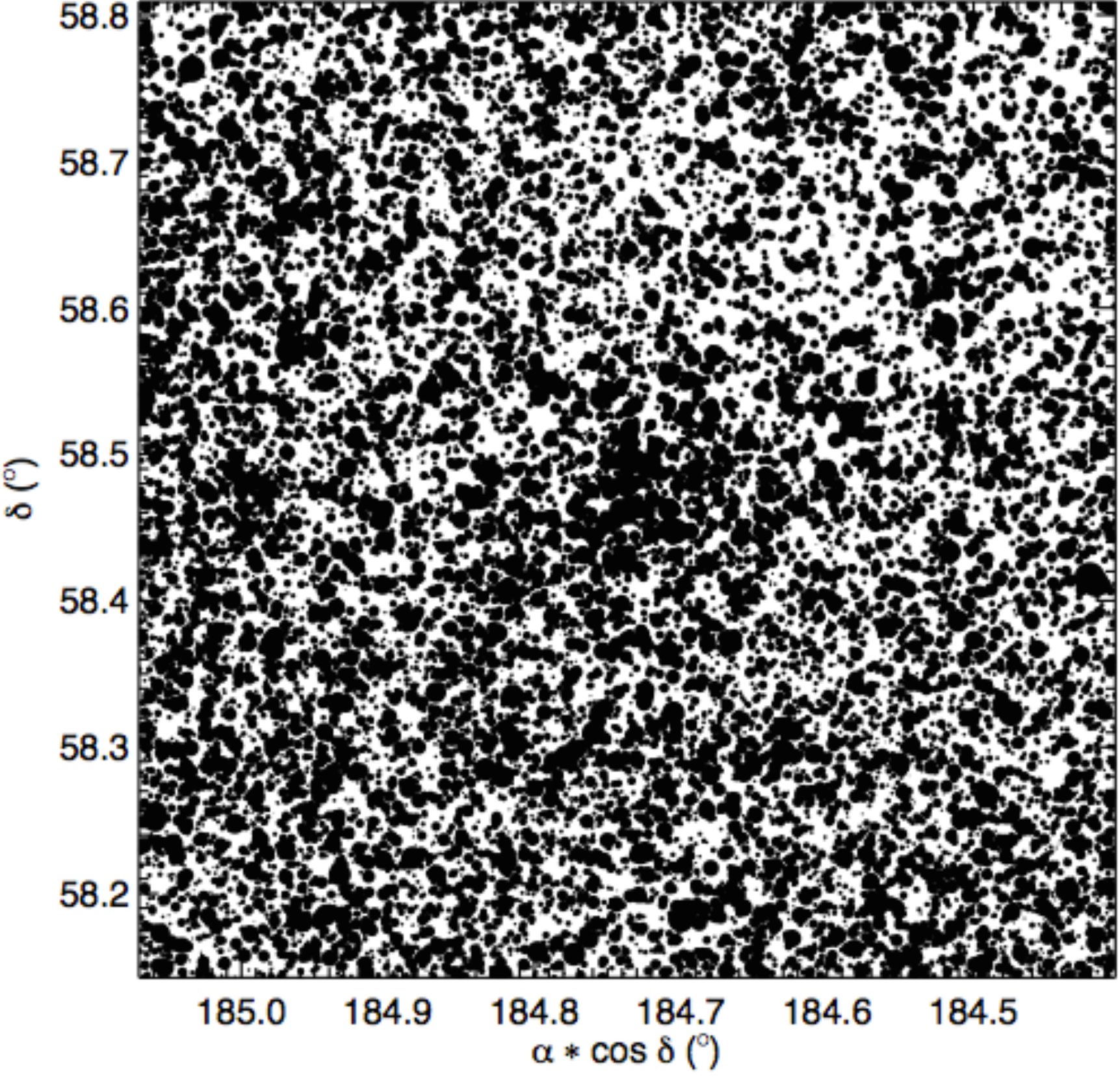} 
    \includegraphics[width=0.45\textwidth]{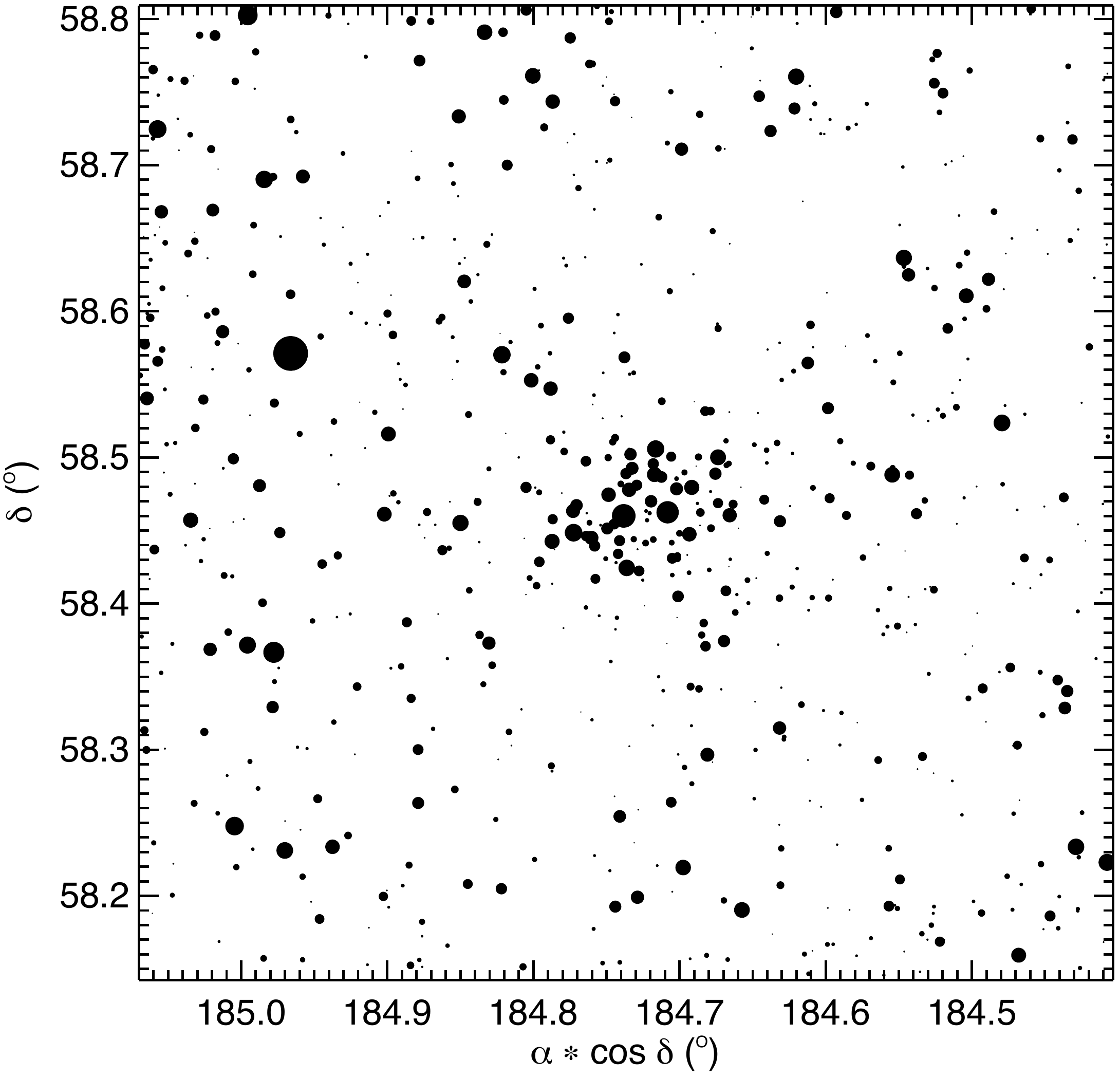}   
    \end{center}
    
  }
\caption{Left panel: skymap for stars in an area of $40{\arcmin}\times40{\arcmin}$ centered on King\,20. Only stars consistent with Arenou et al.'s (2018) filters were kept in our sample. Right panel: skymap after applying the CMD and VPD filters (see text for details).}

\label{King20_skymaps_original_and_filtered}
\end{center}
\end{figure*}
\begin{figure*}
\begin{center}

\parbox[c]{1.0\textwidth}
  {
   
   \begin{center}
    \includegraphics[width=0.265\textwidth]{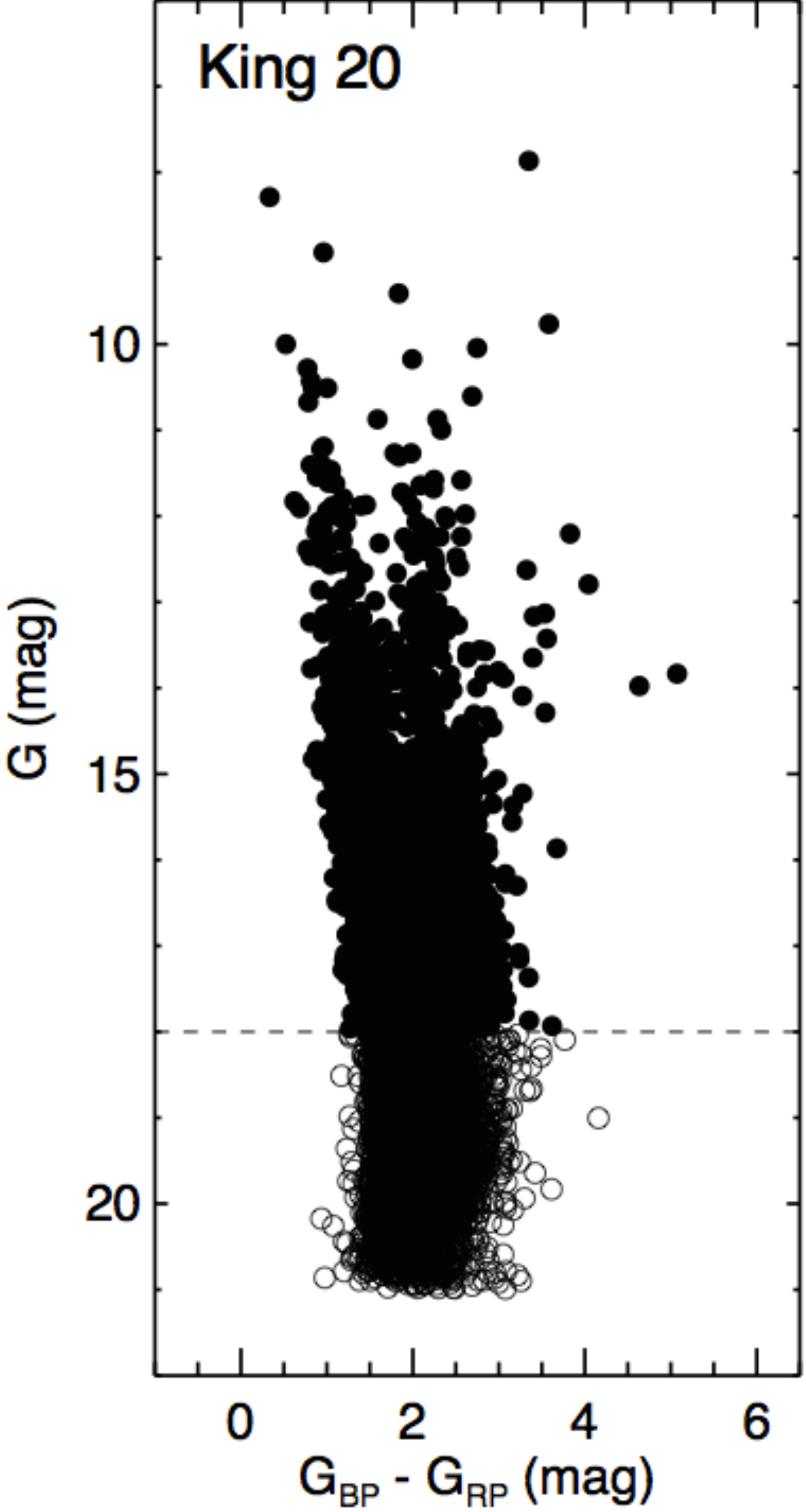}    
    \includegraphics[width=0.513\textwidth]{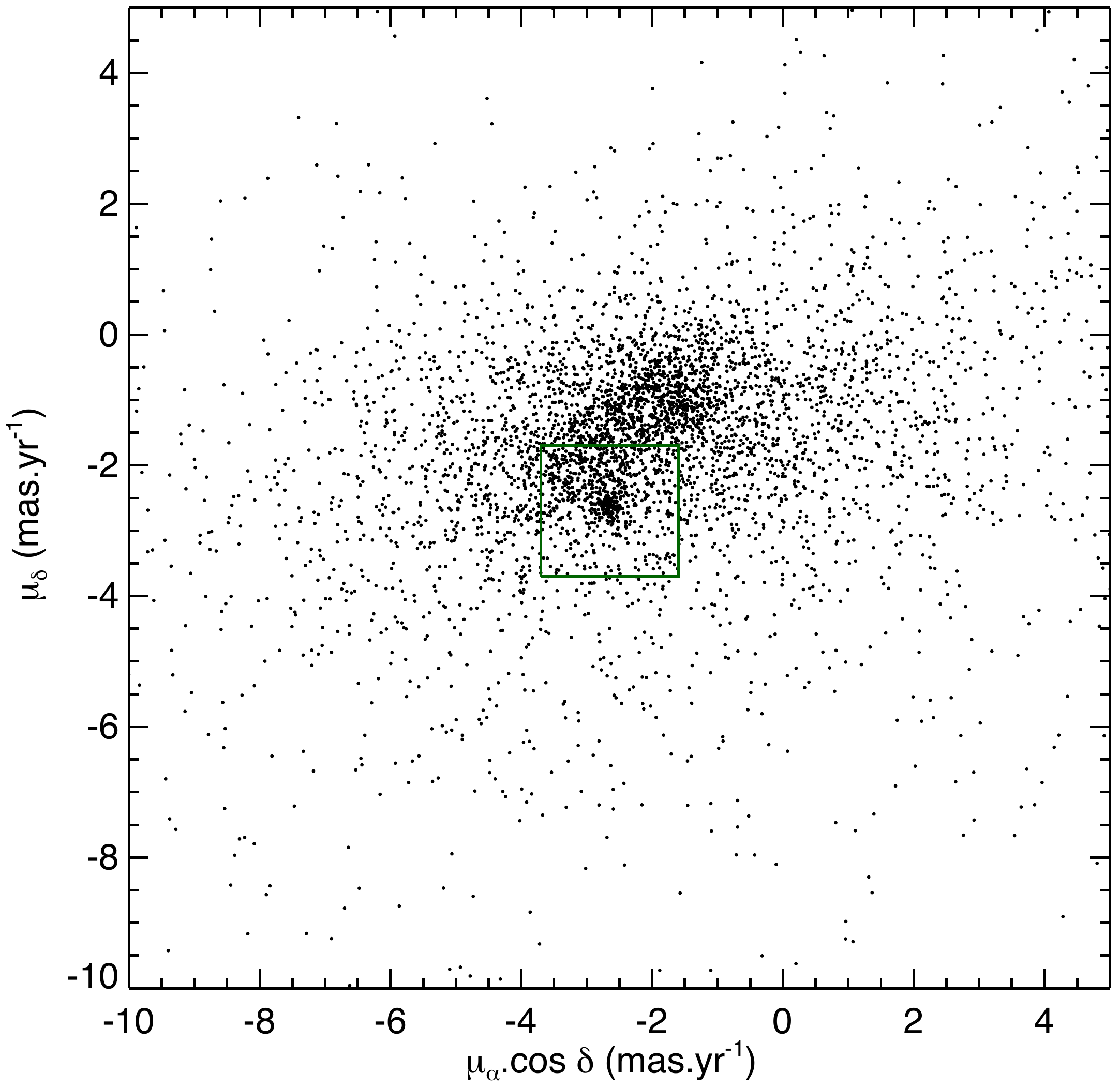}   
    \end{center}
  }
\caption{Left panel: CMD $G\times(G_{\textrm{BP}}-G_{\textrm{RP}})$ for those stars in the skymap of King\,20 shown in Fig.~\ref{King20_skymaps_original_and_filtered}. The dashed line represents the magnitude filter applied to the photometric data. Filled (open) circles represent stars brighter (fainter) than $G=18\,$mag. Right panel: VPD for stars with $G\leq18\,$mag in the region of King\,20. The green square ($2\times2\,$mas\,yr$^{-1}$) shows the VPD filter (see text for details).}

\label{King20_CMD_original_and_VPD_filtered}
\end{center}
\end{figure*}

After applying the basic filtering process, as stated in the previous section, we restricted each cluster data to a squared area with sizes of at least 4 times the limiting radius ($R_{\textrm{lim}}$), as informed in DAML02. The procedure is ilustrated in Fig.~\ref{King20_skymaps_original_and_filtered} for the case of King\,20. We can note a low contrast concentration of bright stars in the central part of its skymap ($40\times40\,$arcmin$^2$ field of view) projected against a dense stellar background ($b\sim-3^{\circ}$). The left panel of Fig.~\ref{King20_CMD_original_and_VPD_filtered} shows the corresponding CMD. We took proper motions for stars with $G\leq18\,$mag (filled circles in the CMD) and built the vector-point diagram (VPD) for them, as shown in the right panel of Fig.~\ref{King20_CMD_original_and_VPD_filtered}.

This filtering process revealed a conspicuous concentration of stars around $(\mu_{\alpha}\textrm{cos}\,\delta, \mu_{\delta})\sim(-2.6, -2.6)$\,mas\,yr$^{-1}$. It would be difficult to verify the existence of such group of stars without the magnitude filter, due to large field contamination mainly by faint disc stars. In order to alleviate the field contamination and thus improve the determination of the cluster's structural parameters (Section \ref{structural_parameters}), we took stars within a squared box with side $2\,$mas\,yr$^{-1}$ centered on the main group (green rectangle in Fig.~\ref{King20_CMD_original_and_VPD_filtered}). This box size is large enough to encompass the member stars without a significant field contamination. The final result of this set of filtering processes is shown in the right panel of Fig.~\ref{King20_skymaps_original_and_filtered}, where the presence of the cluster can be clearly verified. Analogous procedures were employed for the whole sample of investigated OCs. When a clear overdensity in the VPDs could not be obtained, we restricted the cluster's skymap to squared areas of $\sim$2 times the literature $R_{\textrm{lim}}$ and/or applied more restrictive magnitude filters, typically keeping those stars in the range $G\leq17\,$mag.

\subsection{Structural parameters}
\label{structural_parameters}

The second step in our analysis consisted in determining the clusters' central coordinates and structural parameters, namely: core ($r_c$), tidal ($r_t$) and half-mass ($r_{hm}$) radii. To accomplish this, we restricted each cluster to a  subsample of stars consistent with the CMD and VPD filters (that is, magnitude cuts and proper motions box, as defined in the previous section), in order to provide better contrasts between cluster and field populations. To refine the central coordinates, we built a grid of evenly spaced ($\Delta\sim0.5\,$arcmin) $\alpha, \delta$ pairs centered on the literature values (DAML02). At least $\sim$200 pairs were employed during this step. Each $\alpha, \delta$ pair was treated as a tentative centre, for which we built radial density profiles (RDP) by counting the number of stars within concentric annular regions and dividing this number by the ring's area. 

The rings widths were varied from 0.5 to 1.50\,arcmin, in steps of 0.25\,arcmin, and the set of data was superimposed on the same RDP. This strategy was adopted to improve statistics and because the narrower rings are more adequate to sample the central cluster regions, with greater stellar densities, while the larger rings are ideal to probe the external regions, where larger fluctuations induced by field stars become more severe (e.g., \citeauthor{Maia:2010}\,\,\citeyear{Maia:2010}). Uncertainties in star counts come from Poisson statistics. The background levels ($\sigma_{\textrm{bg}}$) and its 1$\sigma$ fluctuation were estimated by taking into account the average and dispersion of the stellar densities corresponding to the rings beyond the limiting radius ($R_{\textrm{lim}}$). $R_{\textrm{lim}}$ is the distance of the cluster centre beyond which the stellar densities fluctuate around a nearly constant value ($\sigma_{\textrm{bg}}$).

\begin{figure*}
\begin{center}

\parbox[c]{1.0\textwidth}
  {
   \begin{center}
    \includegraphics[width=0.95\textwidth]{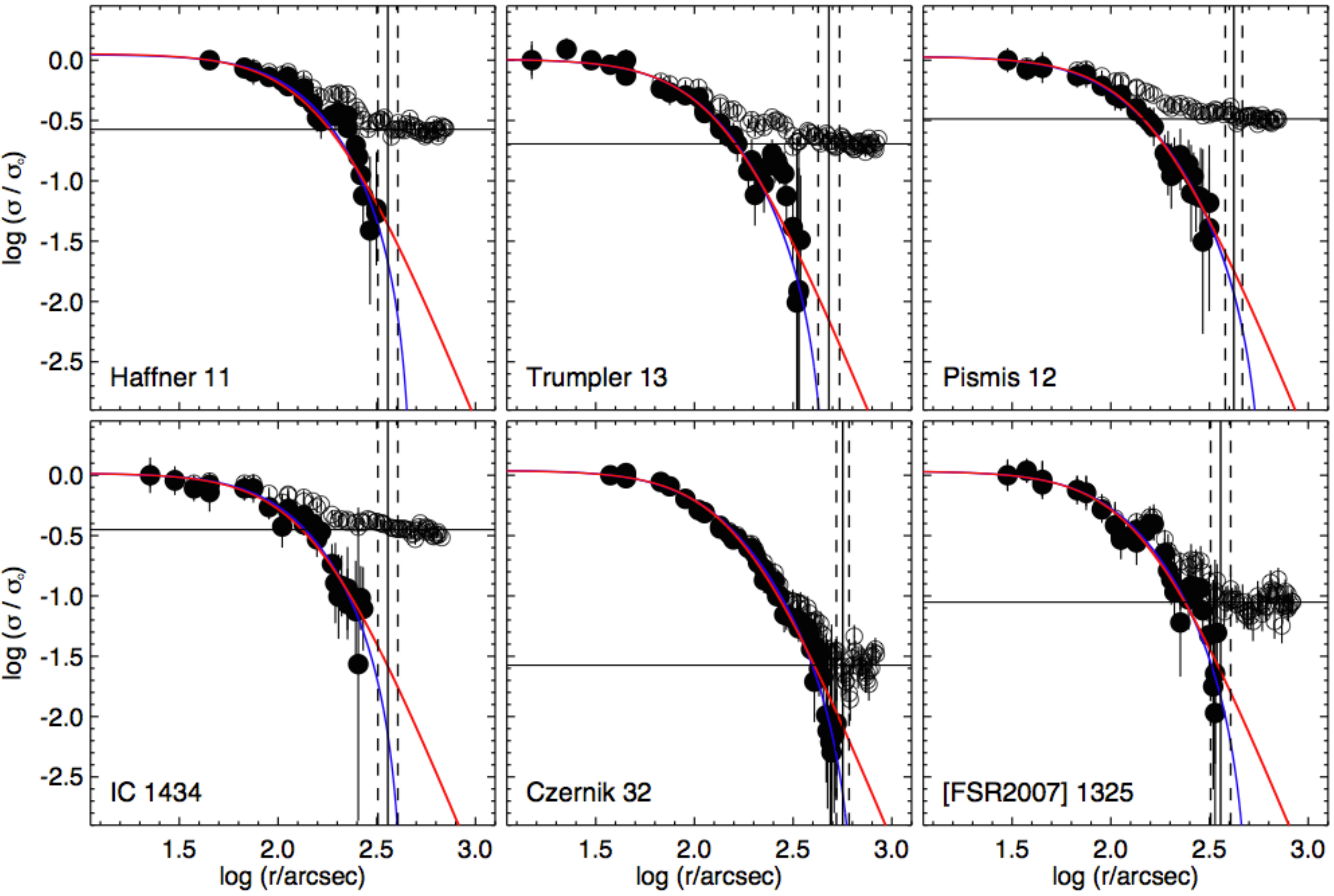} 
    \end{center}
    
  }
\caption{Background subtracted and normalized RDPs (filled symbols) for 6 of the investigated OCs (RDPs for other clusters are shown in the Appendix). The non-subtracted RDPs are plotted with open circles. The blue (red) continuous lines represent the best fitted King (Plummer) profile to our data and the horizontal continuous line represents the mean background density ($\sigma_{\textrm{bg}}$). The vertical continuous and dashed lines represent, respectively, the cluster's limiting radius ($R_{\textrm{lim}}$) and its uncertainty.} 

\label{RDPs_parte1}
\end{center}
\end{figure*}

For each tentative centre, we fitted the corresponding background subtracted RDP using \cite{King:1962}'s model, defined by the expression:

\begin{equation}
   \sigma(r)\,\propto\,\left(\frac{1}{\sqrt{1+(r/r_c)^2}} - \frac{1}{\sqrt{1+(r_t/r_c)^2}}\right)^2
\end{equation}


\noindent
The fits were performed by using a grid of $r_c$ (varying from 0.1${\arcmin}$ to 30${\arcmin}$) and $r_t$ (varying from 0.1${\arcmin}$ to 60${\arcmin}$). After stepping through all coordinates in the grid, we searched for the $\alpha,\,\delta$ pair that resulted in the minimum $\chi^2$ and, at the same time, the highest density in the central cluster region. That is, we searched for central coordinates which resulted in a smooth RDP with considerable stellar overdensity compared to the background. We also performed independent fits of \cite{Plummer:1911} profiles to each cluster RDP

\begin{equation}
   \sigma(r)\,\propto\,\left(\frac{1}{1+(r/a)^2}\right)^2
\end{equation}

\noindent
in order to get estimates of its half-mass radius through the relation: $r_{hm}\sim1.3a$. 

All RDPs were normalized to unity at the innermost radial bins and we summed in quadrature the 1$\sigma$ fluctuation of the background density to each background subtracted bin. The results of our procedure are shown in Fig.~\ref{RDPs_parte1} for six clusters. The complete set of figures for the rest of our sample is shown in the Appendix, which is available online. As can be noted from this figure, both King and Plummer profiles are nearly indistinguishable in the inner regions of each cluster, but differ in the outer parts, where the King profiles have better performance. The results for the structural parameters (already converted to pc using the distance moduli in Table~\ref{fundamental_params}) are listed in Table \ref{struct_params}.

\subsection{Decontamination method}
\label{decontam_method}

The third part of our method is devoted to the systematic search for member stars, that is, disentanglement between cluster and field populations. To accomplish this task, we employed a decontamination algorithm which explores the 3D astrometric space ($\mu_{\alpha}\textrm{cos}\,\delta$, $\mu_{\delta}$, $\varpi$) defined by stars in the cluster region and in a comparison field and establishes membership likelihoods. Our method is fully described in \citeauthor{Angelo:2019a}\,\,(2019a), for which we provide an overview in what follows. 

The main assumption of the method is that stars in a cluster must be more concentrated in a given region of the 3D space compared to field stars. This was the same principle assumed in \citeauthor{Cantat-Gaudin:2018a}\,\,(2018a) and extensively employed in \citeauthor{Cantat-Gaudin:2018b}\,\,(2018b). Firstly, we took photometric and astrometric data for all stars located within the clusters' $r_t$ (Section~\ref{structural_parameters}) and restricted this set of data to those stars inside the box defined in the clusters' VPD (see Section~\ref{preanalysis} and Fig.~\ref{King20_CMD_original_and_VPD_filtered}). From now on, the magnitude filter (as exemplified in the left panel of Fig.~\ref{King20_CMD_original_and_VPD_filtered}) was dismissed and our data were subject only to Arenou et al.'s filters. Stars inside a concentric annular comparison field (internal radius equal to $3r_t$) were also selected following the same filtering process. For all investigated clusters, the area of the comparison field corresponds to the 3 times the cluster area, which allowed a proper sampling of the astrometric space without oversampling.     

Then the astrometric space defined by our data is subdivided in 3D cells (see section 3.2 of \citeauthor{Angelo:2019a}\,\,2019a) and the dispersion of data in both cluster and field samples are statistically compared. A membership likelihood ($l_{\textrm{i}}$) is computed for the i-${th}$ star by using normalized multivariate Gaussian distributions which take into account the three astrometric parameters simultaneously together with the associated uncertainties. The calculation also incorporates the correlations between parameters (equations 1 and 2 of  \citeauthor{Angelo:2019a}\,\,2019a; see also \citeauthor{Dias:2018}\,\,\citeyear{Dias:2018}). The total likelihood of a group of stars is then taken multiplicatively, from which we defined an objective function analogous to the entropy of the parameters space:

\begin{equation}
   S=-\textrm{log}\left( \prod_{i = 1}^{n} l_{i} \right)
,\end{equation}

\noindent
where $n$ is the number of stars within a given 3D cell. The same calculation was performed for both cluster and field samples. 

Those cells for which $S_{\textrm{cluster}}<S_{\textrm{field}}$ were identified and cluster stars within them were flagged (``1") as possible members. For cluster stars within those cells, we additionally computed an exponential factor ($L_{\textrm{star}}$; equation 4 of \citeauthor{Angelo:2019a}\,\,2019a) which evaluates the overdensity of cluster stars within a given cell (flagged as ``1") relatively to the whole grid configuration. This was a necessary procedure to ensure that only significant local overdensities, statistically distinguishable from the distribution of field stars, would receive apreciable membership likelihoods. Finally, cell sizes are increased and decreased by 1/3 of their original sizes in each dimension and the procedure is repeated. The final likelihood for a given star corresponds to the median of the set of $L_{\textrm{star}}$ values taken through the whole grid configurations. 

\begin{figure*}
\begin{center}

\parbox[c]{1.0\textwidth}
  {
   
    \includegraphics[width=0.47\textwidth]{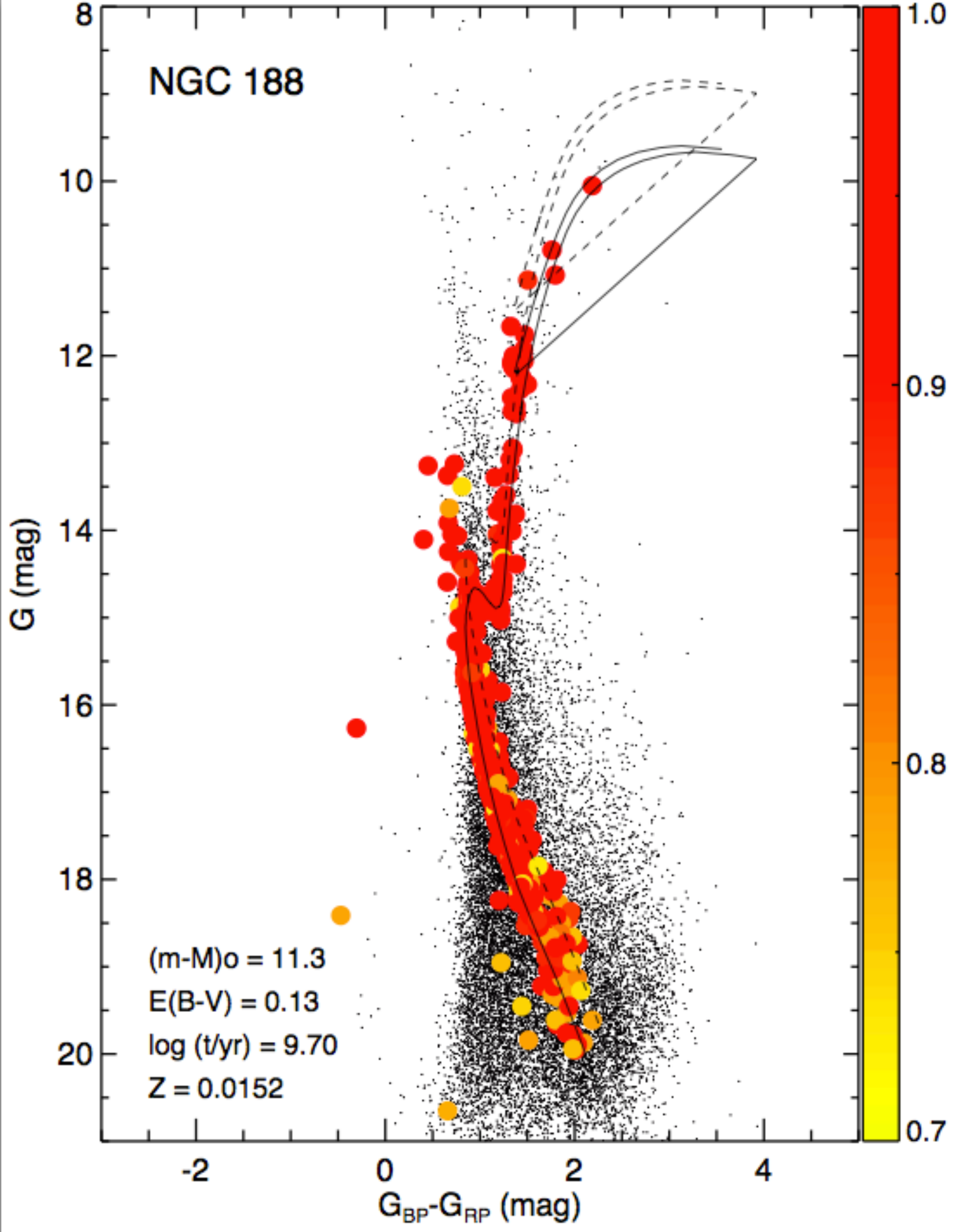}
    \includegraphics[width=0.53\textwidth]{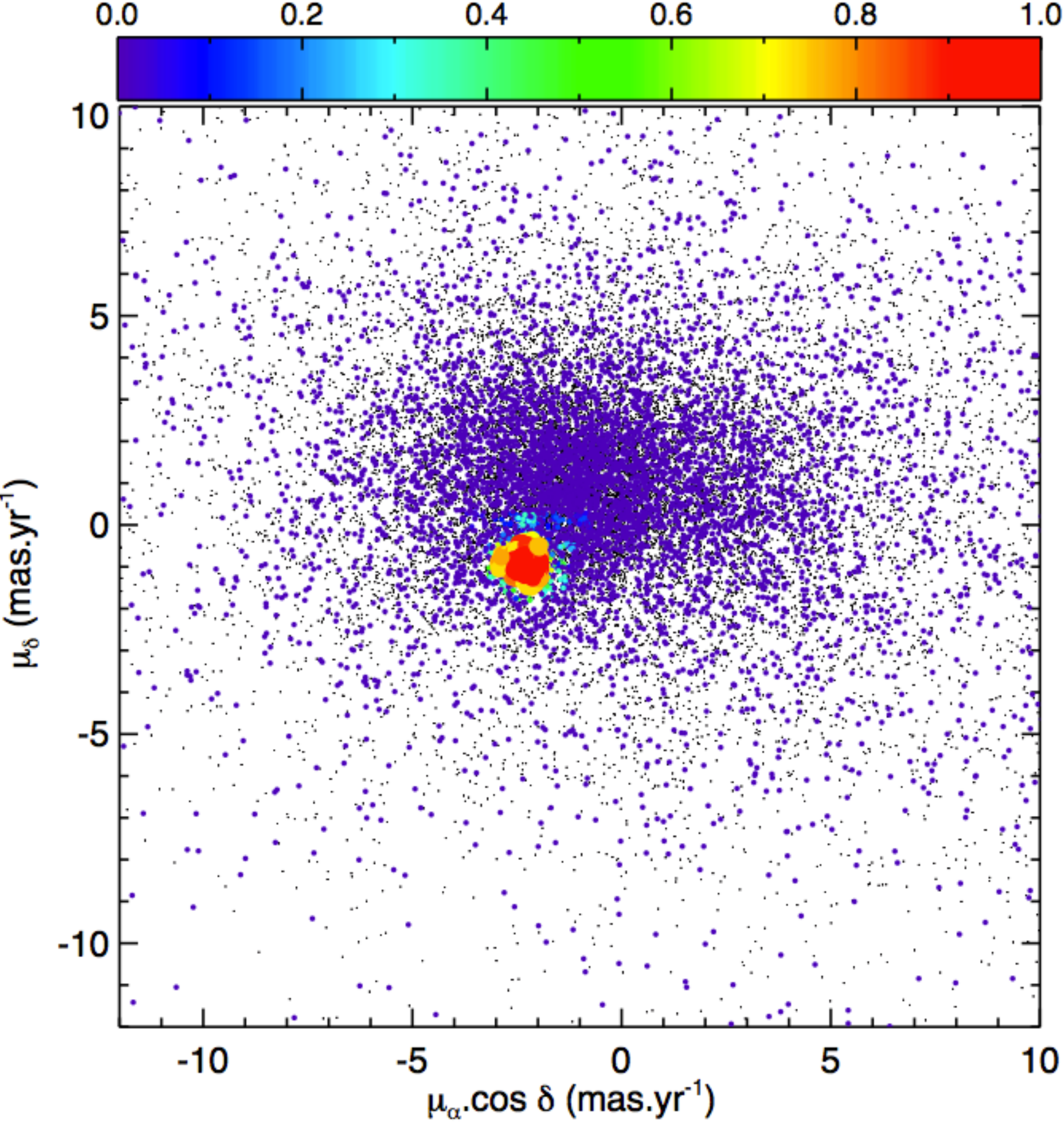} 

    \begin{center}    
        \includegraphics[width=0.50\textwidth]{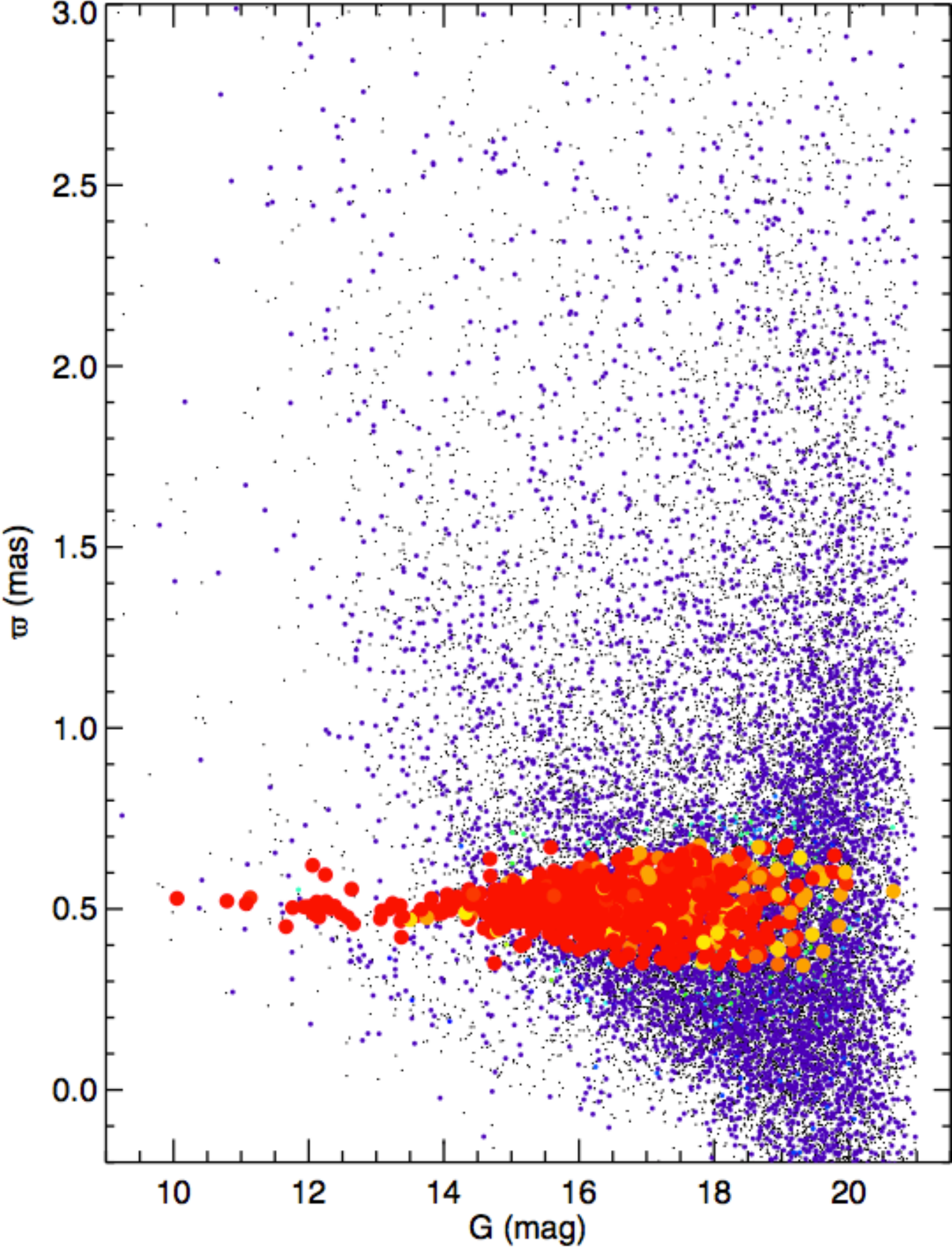}
    \end{center} 
    
  }
\caption{Top-left panel: Decontaminated CMD for NGC\,188. The continuous line is a solar metallicity PARSEC isochrone and the dashed line is the locus of unresolved binaries with same mass components. Top-right panel: cluster's VPD. Bottom panel: Parallax versus $G$ magnitude plot. In all plots, coloured symbols are stars inside the cluster's $r_t$ and colours were attributed according to the membership likelihoods. Big filled circles represent member stars and small black dots represent stars in a comparison field.}

\label{NGC188_CMD_and_VPD_decontam}
\end{center}
\end{figure*}

We illustrate the results of the proposed method by applying it to the well-known OC NGC\,188. In the top-left panel of Figure \ref{NGC188_CMD_and_VPD_decontam}, we can see that the highest membership stars ($L_{\textrm{star}}\gtrsim$\,0.7) define clear evolutionary sequences in the $G\times(G_{\textrm{BP}}-G_{\textrm{RP}})$ CMD and also detached concentrations in both VPD (top-right panel) and $\varpi\,\times\,G$ magnitude plot (bottom panel). Symbol colours were given according to the membership likelihoods, as indicated in the colourbars. Big filled circles represent member stars and small black dots are stars in a comparison field. The continuous line in the decontaminated CMD represents a best-fit log($t\,$yr$^{-1}$)=9.7 solar metallicity PARSEC \citep{Bressan:2012} isochrone (see Section~\ref{results}). The dashed line is the same isochrone displaced vertically by -0.75\,mag and represents the locus of unresolved binary stars with equal mass components.  The fundamental astrophysical parameters are indicated.

\section{Results}
\label{results}

The procedures outlined in Section~\ref{method} were applied to our complete sample (Tables~\ref{struct_params} and \ref{fundamental_params}). The results are shown in the decontaminated CMDs of Fig.~\ref{CMDs_part1} for 6 of our investigated OCs. CMDs for the remaining clusters are shown in the Appendix. We considered only stars with membership likelihoods $L\ge0.7$, since these objects define key evolutionary regions such as the lower main sequence, the turnoff and the red clump, which allowed the estimation of fundamental astrophysical parameters via isochrone fitting, as explained below. All CMDs have been plotted with the same ranges in magnitude and colour indexes to allow a prompt comparison between clusters in terms of magnitude interval, extension of the main sequence and interstellar reddening. 

In some cases (e.g., Haffner\,9 and Lynga\,12) colour filters (analogous to those employed by \citeauthor{Bica:2011}\,\,\citeyear{Bica:2011}) were applied to remove very reddened stars previously to the run of the decontamination method. Those stars are more probably part of the disc population. Proper motions of these  clusters are comparable to the bulk movement of the field (see Appendix), which diminishes the contrast between both populations in the astrometric space, increasing the presence of outliers in the final solution. Applying colour filters was useful to reduce the residual contamination, thus resulting in clearer evolutionary sequences in the corresponding CMDs.

\begin{figure*}
\begin{center}

\parbox[c]{0.99\textwidth}
  {
   \begin{center}
     \includegraphics[width=0.99\textwidth]{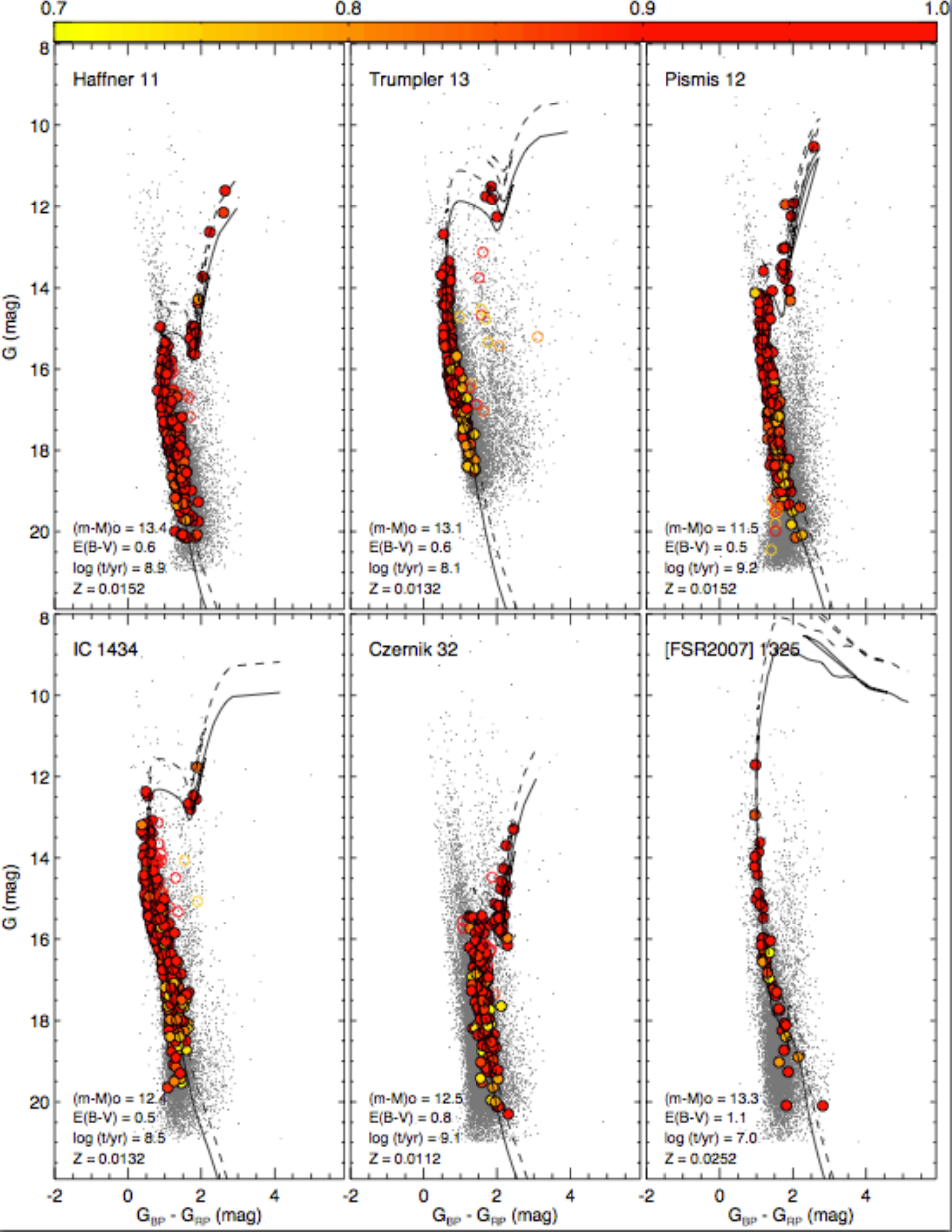}    
    \end{center}
    
  }
\caption{CMDs $G\times(G_{\textrm{BP}}-G_{\textrm{RP}})$ for the OCs Haffner\,11, Trumpler\,13, Pismis\,12, IC\,1434, Czernik\,32 and [FSR2007]\,1325. The symbols convention is the same of those adopted in Fig.~\ref{NGC188_CMD_and_VPD_decontam}. Only stars with $L\ge0.7$ are shown. Member stars are plotted as filled circles. Best-fit PARSEC isochrones (continuous lines) and equal-mass binary locii (dashed lines) are also indicated.}

\label{CMDs_part1}
\end{center}
\end{figure*}

\subsection{Isochrone fitting}
\label{isoc_fitting}

The fundamental astrophysical parameters age, distance modulus, reddening and metallicity (respectively, log\,$t$, $(m-M)_0$, $E(B-V)$ and $[Fe/H]$) were derived using the automated isochrone fitting module of ASteCA\footnote[2]{Available at: https://asteca.github.io/} (Automated Stellar Cluster Analysis) code. The method is fully presented and tested in section 2.9 of \cite{Perren:2015}. Briefly, it consists on the generation of synthetic clusters from theoretical isochrones with a variety of parameters and the selection of the best fit through a genetic algorithm.  

Previously to the run of the automated method, we performed visual isochrone fits in order to obtain reasonable initial guesses (log\,$t_{\textrm{ini}}$, $(m-M)_{0,\textrm{ini}}$, $E(B-V)_{\textrm{ini}}$ and $[Fe/H]_{\textrm{ini}}$) for the fundamental parameters. For each cluster, the initial guess for the distance modulus was based on the mean parallax ($\langle\varpi\rangle$) of high membership stars ($L\ge0.7$; Fig.~\ref{CMDs_part1}) through the relation $(m-M)_0=5\,\textrm{log}\,(100/\langle\varpi\rangle)$, with $\langle\varpi\rangle$ expressed in mas, as usual. $E(B-V)_{\textrm{ini}}$ was taken from DAML02 catalogue.  When necessary, some adjustments were applied to both parameters in order to obtain a reasonable fit to the locus of points along the main sequence. At this stage, solar metallicity isochrones were employed. After that, log\,$t_{\textrm{ini}}$ and $[Fe/H]_{\textrm{ini}}$ were estimated from brighter stars along the more evolved sequences. The relative distance between the cluster turnoff and the red clump (if present) is a very useful constraint for isochrone fitting.   

Taken the initial guesses for the fundamental parameters, we then restricted the parameters space covered by the PARSEC models and, as input to the ASteCA code, the intervals were sampled according to the following specifications:

\begin{itemize}
     \item $(m-M)_{0,\textrm{ini}}$\,-\,0.5\,$\leq$\,$(m-M)_0\leq(m-M)_{0,\textrm{ini}}\,+\,0.5$, in steps of $\Delta(m-M)_0=0.1\,$mag; 
     \item $E(B-V)_{\textrm{ini}}-0.5\leq E(B-V)\leq E(B-V)_{\textrm{ini}}+0.5$, in steps of $\Delta E(B-V)=0.02\,$mag; 
          \item $Z_{\textrm{ini}}-0.01\leq Z\leq Z_{\textrm{ini}}+0.01$, in steps of $\Delta Z=0.002$,      
\end{itemize}

\noindent where the overall metallicity $Z$ can be associated to the iron abundance ratio $[Fe/H]$ according to the approximate relation $[Fe/H]\sim\textrm{log}\,(Z/Z_{\odot})$ \citep{Bonfanti:2016}, with $Z_{\odot}=0.0152$ \citep{Bressan:2012}. Besides the intervals specified above, ASteCA builds synthetic clusters with total masses $M$ between 10$-$10000\,$M_{\odot}$ (steps of $\Delta M=100\,M_{\odot}$) and containing fractions of binary stars ($f_{bin}$) varying from 0 to 1 (steps of $\Delta f_{bin}=0.2$), with mass ratio between the primary and secondary stars fixed in a pre-defined value of 0.7. No magnitude cuts have been applied during the isochrone fitting process. With the above intervals, $\sim10^6$ models were compared to the observed CMD for each studied OC. The CMDs in Fig.~\ref{CMDs_part1}, besides those in the Appendix, show that the method provided adequate fits, with the corresponding isochrones properly describing the evolutionary sequences defined by the more probable member stars.

\begin{figure*}
\begin{center}

\parbox[c]{1.0\textwidth}
  {
   \begin{center}
    \includegraphics[width=1.00\textwidth]{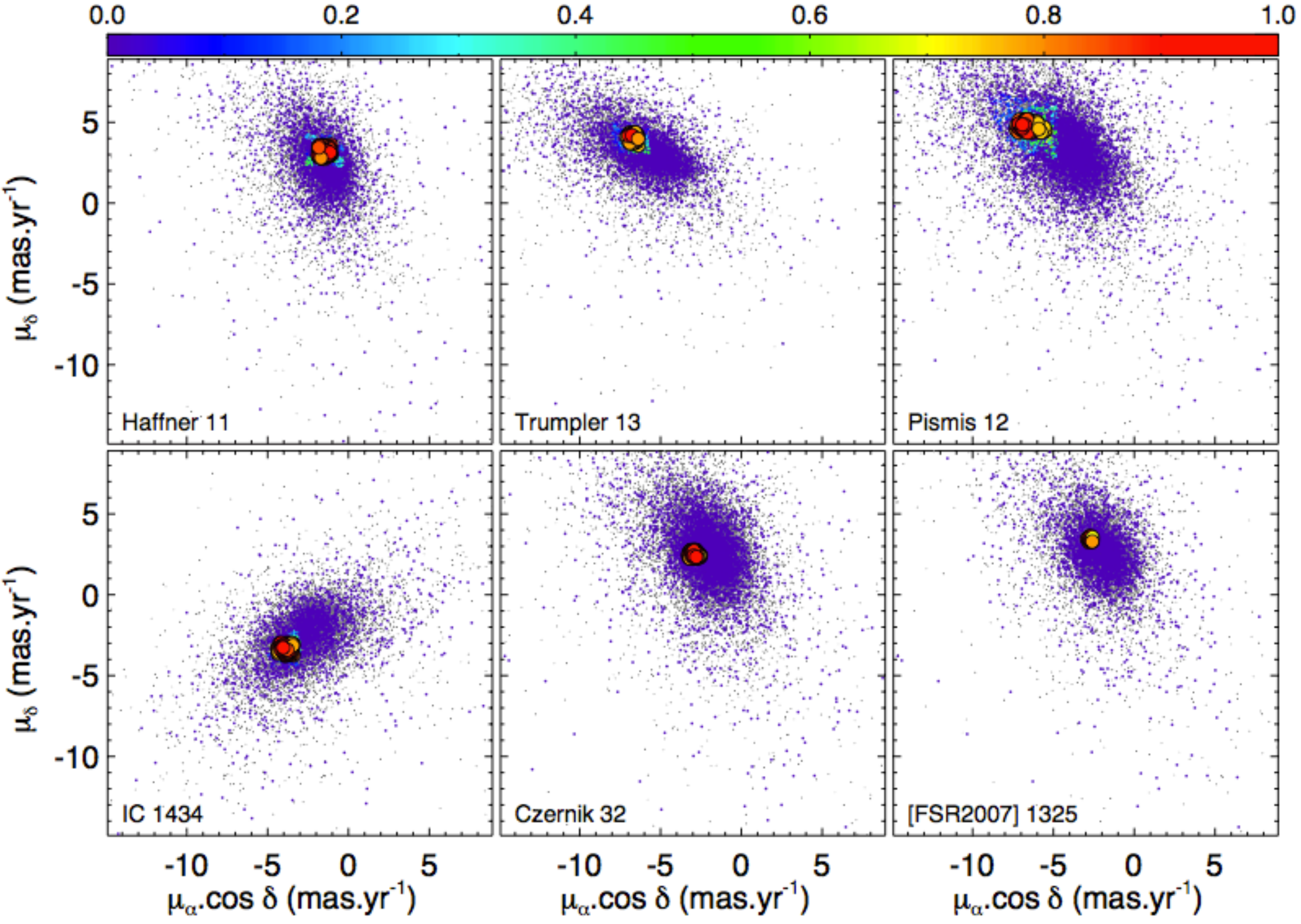}    
    \end{center}
    
  }
\caption{  VPDs for 6 of our studied OCs. The symbols convention is the same of those adopted in Fig.~\ref{NGC188_CMD_and_VPD_decontam}.}

\label{VPDs_part1}
\end{center}
\end{figure*}

\begin{figure*}
\begin{center}

\parbox[c]{1.0\textwidth}
  {
   \begin{center}
    \includegraphics[width=1.00\textwidth]{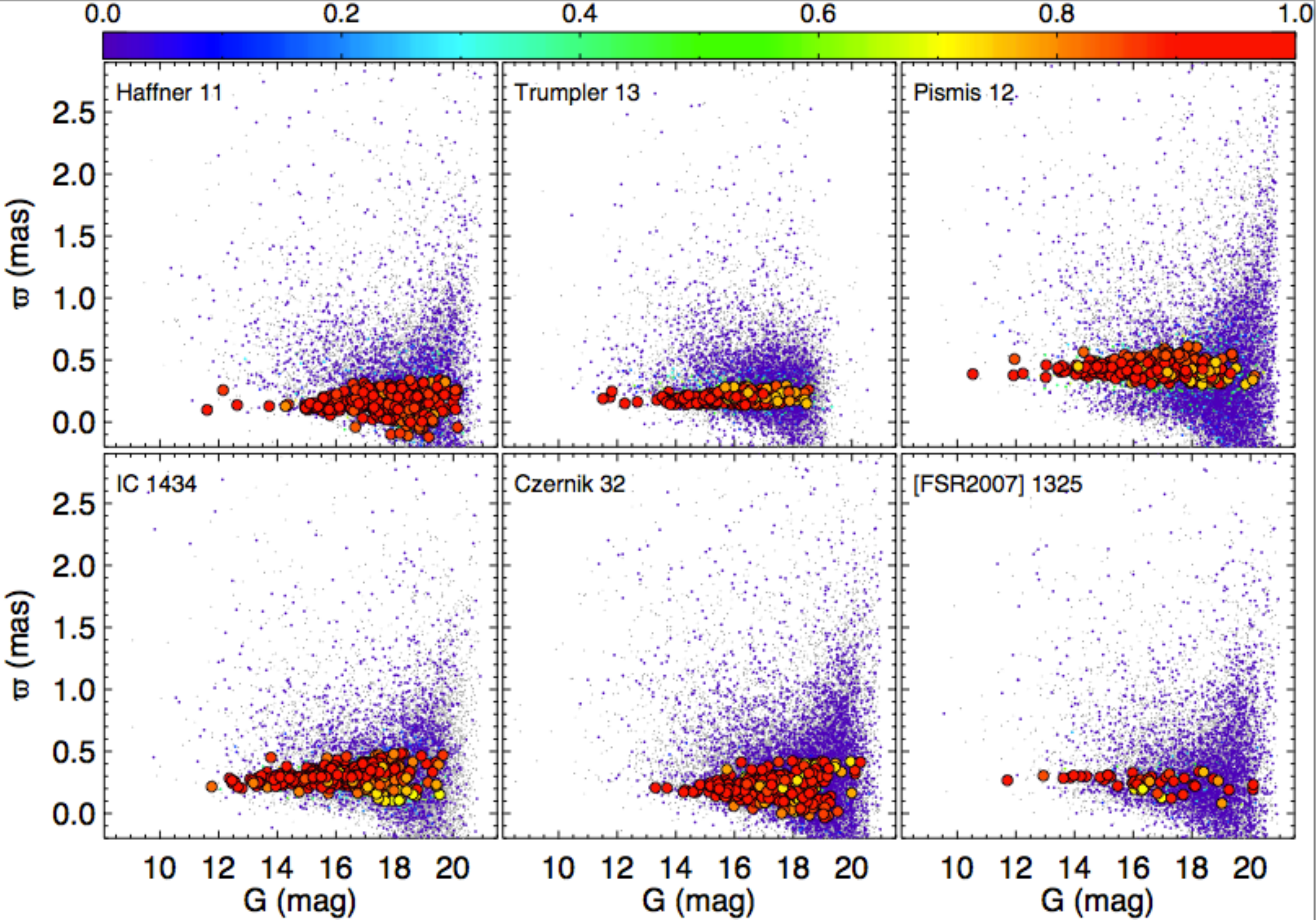}   
    \end{center}
    
  }
\caption{$\varpi\times\,G$\,magnitude for 6 of our investigated OCs. Symbols conventions are the same of those employed in Fig.~\ref{NGC188_CMD_and_VPD_decontam}.}

\label{plx_vs_G_part1}
\end{center}
\end{figure*}

Figs.~\ref{VPDs_part1} and \ref{plx_vs_G_part1} exhibit, respectively, the clusters VPDs and the $\varpi\times\,G$ magnitude plots after applying our decontamination method. The same symbols convention of Fig.~\ref{NGC188_CMD_and_VPD_decontam} were employed here. The corresponding figures for the rest of our sample are shown in the Appendix.

\subsection{Comparison with previous works}
\label{comparison_previous_works}

\citeauthor{Cantat-Gaudin:2018b}\,\,(2018b) performed a uniform characterization of 1212 Galactic clusters by applying an unsupervised membership assignment code to the Gaia DR2 data contained within the fields of those clusters. Their decontamination method is based on the same strategy of the present work, which is to look for concentrations of cluster stars in the astrometric space that are more tightly distributed than field stars. A basic difference is that \citeauthor{Cantat-Gaudin:2018a}\,\,(2018a)'s method performs comparisons between the distribution of astrometric data for cluster stars with that for randomly generated samples. In our case, we have employed \textit{observed} samples of stars in comparison fields (Section~\ref{decontam_method}). Besides, the sampling of the astrometric space in \citeauthor{Cantat-Gaudin:2018a}\,\,(2018a) is based on the $k$-means clustering algorithm incorporated to the membership assignment method UPMASK \citep{Krone-Martins:2014}. In our case, the astrometric space is sampled using  uniform grid of cells with varying sizes (Section \ref{decontam_method}). 

\begin{figure*}
\begin{center}

\parbox[c]{1.0\textwidth}
  {
   \begin{center}
    \includegraphics[width=1.00\textwidth]{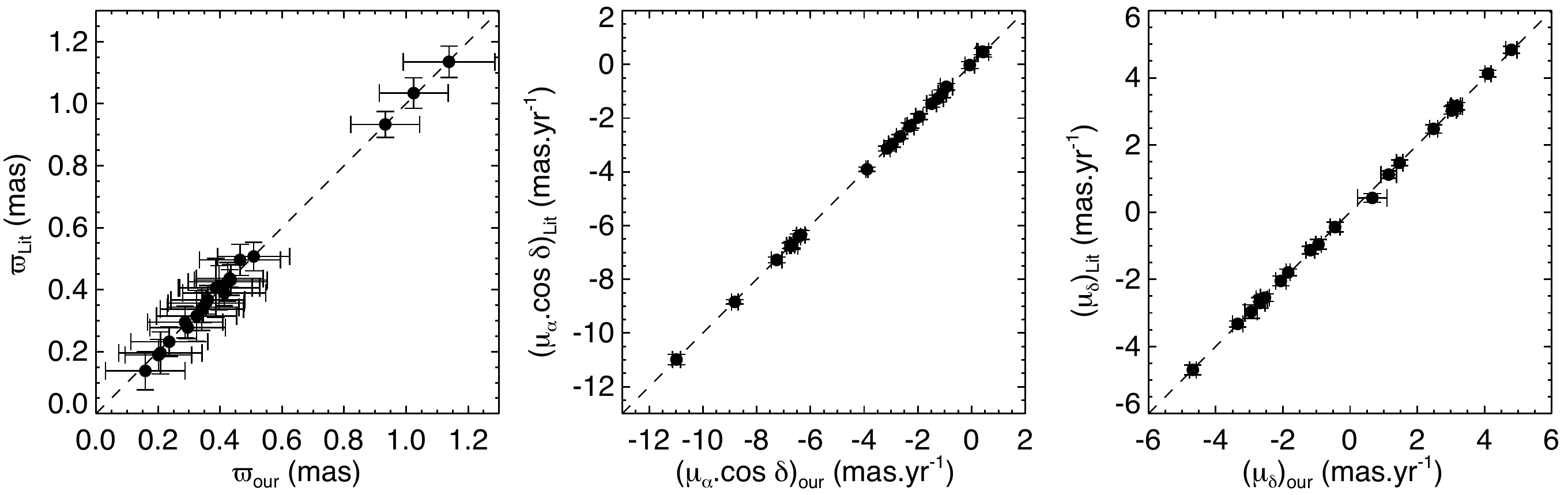}    
    \end{center}
    
  }
\caption{Comparison between mean clusters' parameters derived in the present study ($\langle\mu_{\alpha}\textrm{cos}\,\delta\rangle_{\textrm{our}}$, $\langle\mu_{\delta}\rangle_{\textrm{our}}$, $\langle\varpi\rangle_{\textrm{our}}$) and those obtained by Cantat-Gaudin et al. (2018b) ($\langle\mu_{\alpha}\textrm{cos}\,\delta\rangle_{\textrm{lit}}$, $\langle\mu_{\delta}\rangle_{\textrm{lit}}$, $\langle\varpi\rangle_{\textrm{lit}}$). In our sample, error bars correspond to the intrinsic dispersion $(\sigma_{\mu_{\alpha}}$, $\sigma_{\mu_{\delta}}$, $\sigma_{\varpi}$) obtained from each sample of cluster's members (for $\sigma_{\varpi}$, we have added in quadrature a systematic uncertainty of 0.1\,mas). The dashed line is the identity relation.}

\label{compara_com_Cantat_Gaudin}
\end{center}
\end{figure*}

We performed a direct comparison between the mean astrometric parameters ($\langle\mu_{\alpha}\textrm{cos}\,\delta\rangle$, $\langle\mu_{\delta}\rangle$, $\langle\varpi\rangle$) derived for OCs in the present paper and the corresponding ones obtained by \citeauthor{Cantat-Gaudin:2018b}\,\,(2018b). Fig.~\ref{compara_com_Cantat_Gaudin} compares the results for 22 OCs in common with both studies (data for Collinder\,351, [FSR2007]\,1325, Herschel\,1, Lynga\,12 and Ruprecht\,30 are absent in \citeauthor{Cantat-Gaudin:2018b}\,\,2018b). We can see that, despite their differences, both methods recovered almost the same mean parameters. In our case, the error bars in Fig.~\ref{compara_com_Cantat_Gaudin} correspond to the intrinsic dispersion of the astrometric parameters considering each sample of member stars (Sections~\ref{method} and \ref{results}). Concerning the parallaxes, to the final dispersions we have added in quadrature a systematic uncertainty of 0.1\,mas affecting the astrometric solution in Gaia DR2 \citep{Luri:2018}.

\begin{figure*}
\begin{center}

\parbox[c]{1.0\textwidth}
  {
   \begin{center}
    \includegraphics[width=1.00\textwidth]{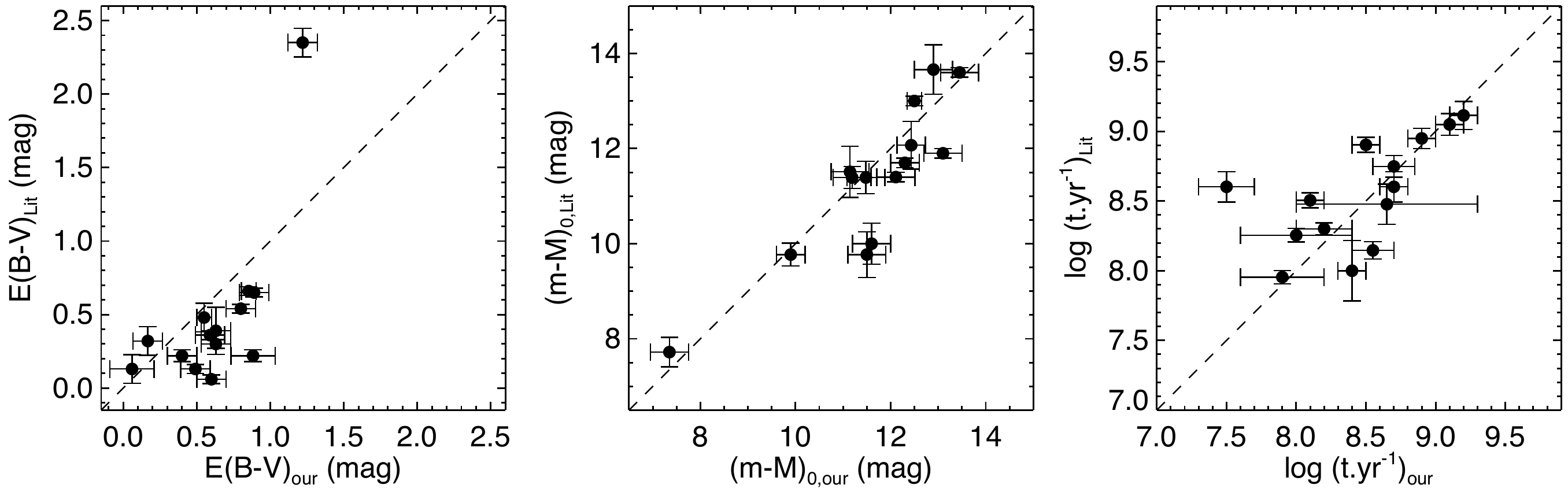}    
    \end{center}
    
  }
\caption{Comparison between the fundamental parameters $E(B-V)$, $(m-M)_0$ and log\,$t$ derived in the present study and those obtained in the literature. The dashed line is the identity relation.}

\label{compara_com_Bica_e_Bonatto}
\end{center}
\end{figure*}

Fourteen OCs in our main sample (Tables~\ref{struct_params} and \ref{fundamental_params}) were also investigated by \cite{Bica:2005a}, \cite{Bica:2006} and \cite{Bica:2011}. The OCs NGC\,2421 and [FSR2007]\,1325 were included  since they are projected in the same regions of Czernik\,31 and Czernik\,32, respectively (see Section \ref{comments_individual_clusters}). Fig.~\ref{compara_com_Bica_e_Bonatto} shows comparisons between the fundamental parameters $E(B-V)$, $(m-M)_0$ and log\,$(t)$ derived in the present study and the literature ones. As we can see, there is not a tight agreement between both sets of parameters and some severe discrepancies are present. The main reason for such discrepancies, besides the use of different sets of isochrones and CMD fitting techniques, may be related to the CMDs decontamination procedures and member stars selection. The procedures employed in the serie of papers of Bica \& Bonatto are based on the use of $JHK_s$ photometry from the 2MASS catalogue and on the analysis of $J\times(J-H)$ and $J\times(J-K_s)$ CMDs built for stars in the clusters regions and for comparison fields. 

Briefly, their method consists in a decontamination algorithm \citep{Bonatto:2007} which builds 3D CMDs ($J$, $(J-H)$, $(J-K_s)$ axes) divided in cells with different grid configurations and assigns memberhip probabilities according to the local overdensity of stars in the cluster CMD compared to the distribution of field stars in the same data domain in the field CMD. Keeping those stars with higher memberships allowed the construction of decontaminated CMDs. Residual contamination, specially in the case of very reddened clusters projected close to the Galactic plane, is eliminated with the use of colour filters (see, e.g., figures 1 to 3 of \citeauthor{Bica:2011}\,\,\citeyear{Bica:2011}). Then visual isochrone fits were performed and the fundamental astrophysical parameters were derived. 

Our method has the advantage of relying not only on photometric data but also on high precision astrometric information. Specially in the case of low contrast OCs, we advocate that different kind of data should be combined in a joint analysis for a more rigorous members selection. In general, our CMDs are typically $\sim1-5\,$mag deeper than those built with 2MASS filters, which allows better constraints for isochrone fitting. Besides, the parameters obtained in Bica \& Bonatto's papers assumed fixed solar values for the clusters metallicities, for simplicity.

\subsection{Comments on individual clusters}
\label{comments_individual_clusters}

In this section, we highlight some comments about two binary cluster candidates (namely, Czernik\,31$-$NGC\,2421 and Czernik\,32$-$[FSR2007]\,1325) and also for other three OCs (namely, Ruprecht\,30, Ruprecht\,130 and NGC\,3519) which were classified as asterisms or had their real physical nature inconclusively determined in previous studies.

\subsubsection{Czernik\,31 and NGC\,2421}

\begin{figure}
\begin{center}
 \includegraphics[width=8.5cm]{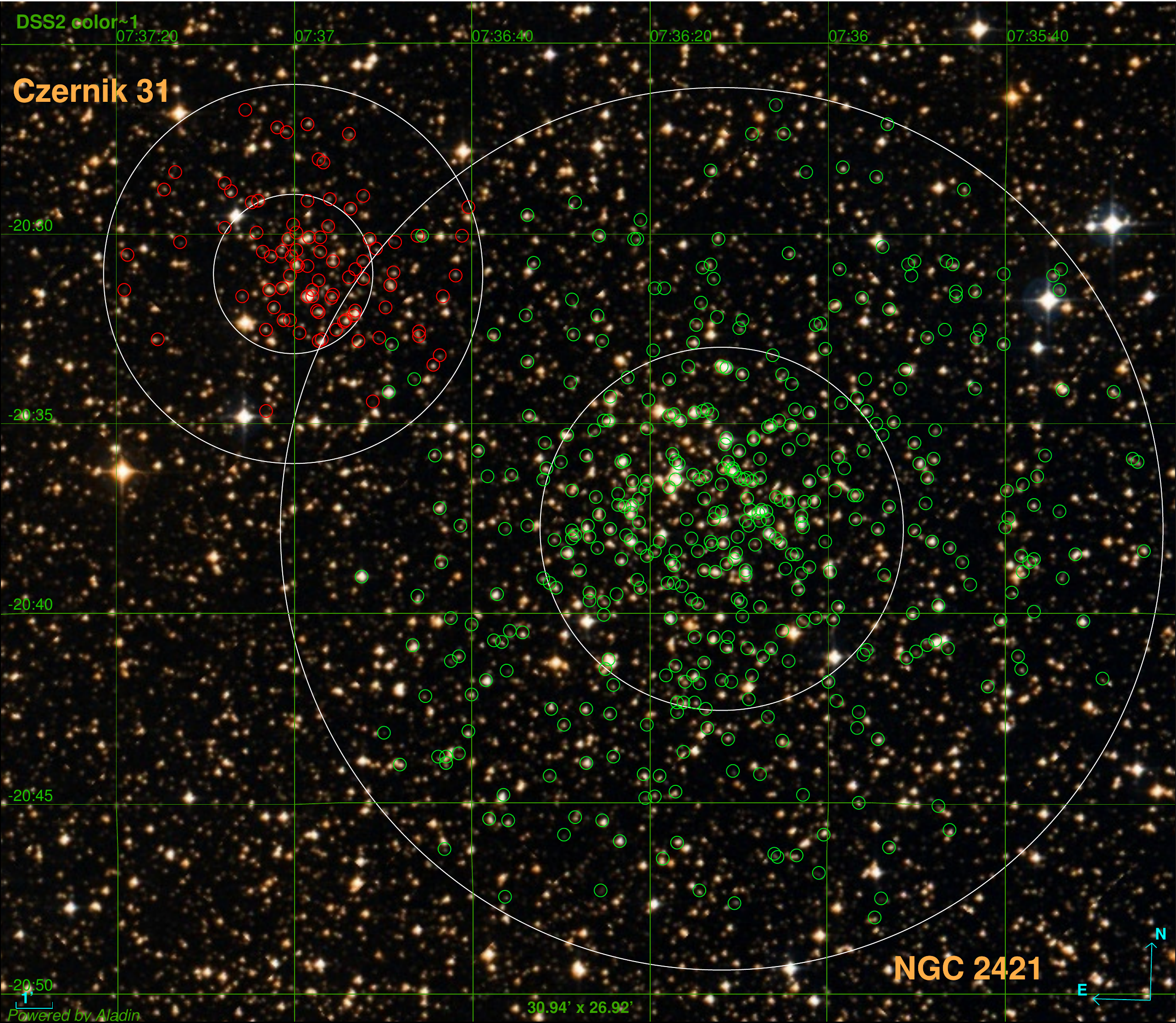}
 \caption{ DSS image (31$\arcmin\times27\arcmin$) of the pair Czernik\,31$-$NGC\,2421. Member stars of each cluster are identified with coloured circles. Concentric white circles represent $r_c$ and $r_t$ radii for each cluster. North is up and East to the left.}
   \label{pair_Czernik31_NGC2421}
   \end{center}
\end{figure}

The pair Czernik\,31$-$NGC\,2421 is present in the list of candidate binary clusters of \cite{de-La-Fuente-Marcos:2009}. Fig.~\ref{pair_Czernik31_NGC2421} shows both clusters with member stars highlighted. Not only are they projected in the same region, but they seem to form a physical pair, since their distance moduli, ages and colour excesses are compatible with each other within uncertainties (Table \ref{fundamental_params}). Besides, the position of the centroids in their corresponding VPDs are quite similar (see Appendix).

NGC\,2421 presents a fluctuation in its RDP between $6.5\arcmin\lesssim r\lesssim8.5\arcmin$ ($2.6\lesssim$\,log(r/arcsec)\,$\lesssim2.7$; see Appendix), where members stars with $12.5\leq G(\textrm{mag})\leq19.8$ (stellar masses in the range $0.70\lesssim M(M_{\odot})\lesssim4.9$) are found. Less noticeably, Czernik\,31 presents a fluctuation in stellar density in the external region $3.5\arcmin\lesssim r\lesssim4.5\arcmin$ ($2.3\lesssim$\,log(r/arcsec)\,$\lesssim2.4$), where we find member stars with $G$ magnitudes in the range $15.3\leq G(\textrm{mag})\leq19.8$ (which converts to stellar masses in the range $0.85\lesssim M(M_{\odot})\lesssim2.2$, considering the cluster distance modulus, metallicity and age; see Section~\ref{results} and Table~\ref{fundamental_params}). Both density fluctuations define these clusters' corona and these external structures may be the result of tidal interactions between them. The overlap of their tidal radii, as seen in Fig.~\ref{pair_Czernik31_NGC2421}, reinforces the hypothesis of close gravitational interaction. Following \cite{Innanen:1972}, for two clusters separated by a distance smaller than three times the outer radius of each cluster, the amount of mutual disruption is not negligible. In fact, the dynamics of interaction between Czernik\,31$-$NGC\,2421 could be investigated in more detail with the determination of radial velocities, in order to obtain 3D velocity vectors.

\subsubsection{Czernik\,32 and [FSR2007]\,1325}

\begin{figure}
\begin{center}
 \includegraphics[width=8.5cm]{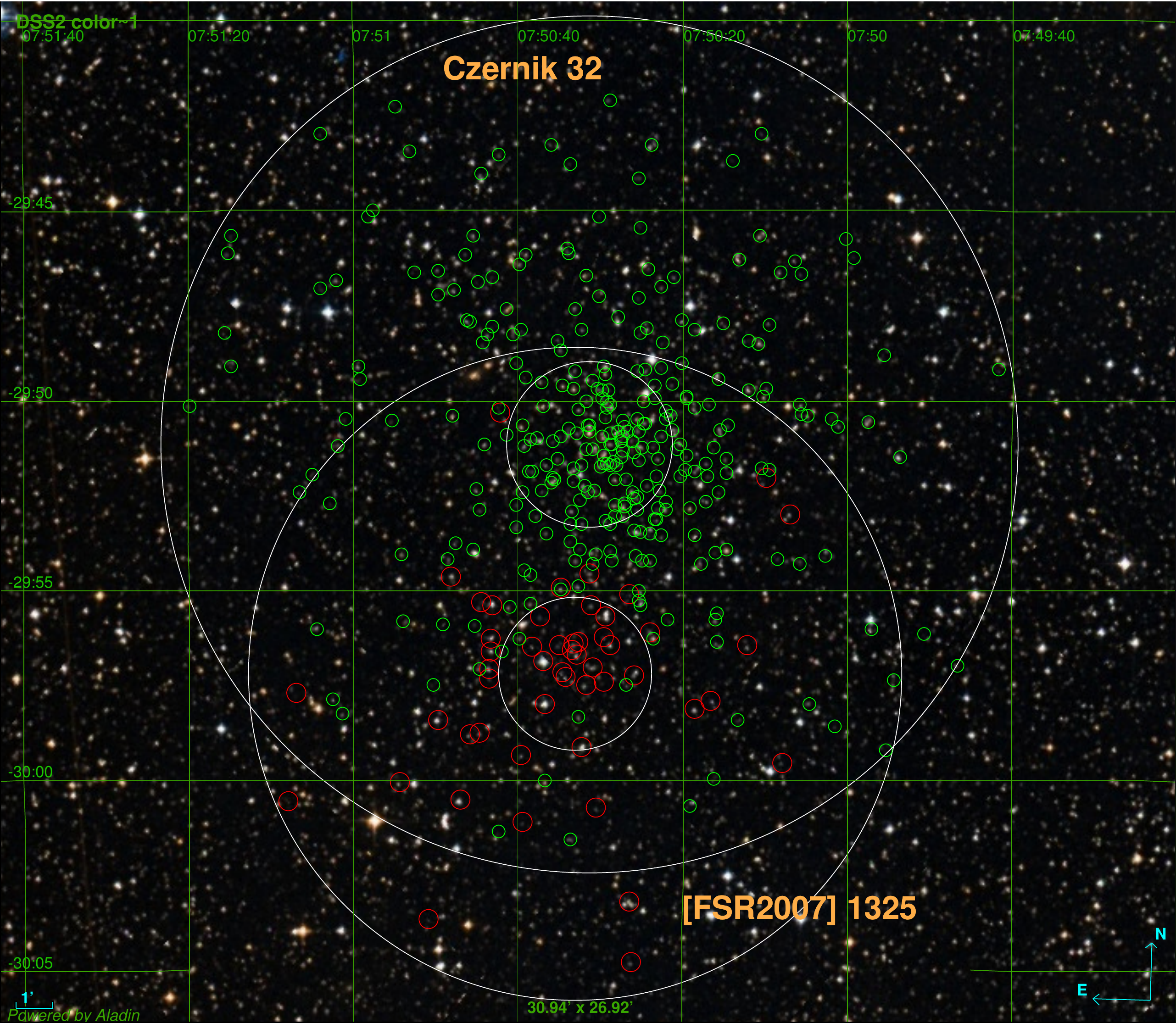}
 \caption{ Same of Fig.~\ref{pair_Czernik31_NGC2421}, but showing the pair Czernik\,32$-$[FSR2007]\,1325. Image size is 31$\arcmin\times27\arcmin$. }
   \label{pair_Czernik32_Czernik32b}
   \end{center}
\end{figure}

The OC [FSR2007]\,1325 is listed in the survey of \cite{Froebrich:2007}, who used star density maps constructed from the 2MASS catalog to locate a sample of star clusters at Galactic latitudes $\vert b\vert<20^{\circ}$. It is projected southwards of Czernik\,32, as shown in Fig.~\ref{pair_Czernik32_Czernik32b}. 

The results in Table~\ref{fundamental_params} show that these two clusters present distances that are incompatible with the hypothesis of a pair of interacting OCs. The corresponding colour excesses are also different. Besides, their RDPs (Fig.~\ref{RDPs_parte1}) are well fitted by King's profile until $r\sim r_t$, do not exhibiting any noticeable distorsions that could be the result of tidal interactions. Therefore, we advocate that the apparent proximity of Czernik\,32 and [FSR2007]\,1325 along the line of sight may be a projection effect.

\subsubsection{Ruprecht\,30 and Ruprecht\,130}

\cite{Carraro:2010} performed star counts in the direction of Ruprecht\,30 using deep photometry in the $V$ filter (down to $V\sim20\,$mag) and did not verify any obvious overdensities (their figure 4), thus suggesting that there is no cluster at the location of Ruprecht\,30. In fact, establishing its physical nature is not a trivial task, since it is not a heavily-populated stellar concentration and it is projected against a dense background. During the pre-analysis procedure (Section~\ref{preanalysis}), we could only find some clues about its physical existence after applying magnitude filters, like that illustrated in Fig.~\ref{King20_CMD_original_and_VPD_filtered}. Star counts on unfiltered data can mask the physical existence of this cluster, due to the prevalence of the field population. After running our decontamination method, we concluded that it \textit{is} a genuine stellar aggregate, which was the same conclusion of \cite{Bica:2011}.

Ruprecht\,130 was included in the list of ``borderline cases"\, as stated by \cite{Bica:2011}, since no definitive conclusions could be drawn regarding its physical nature based on decontaminated $J\times(J-K_s)$ CMD (their figure 3). They suggest the presence of a real cluster, which is confirmed by our analysis. Ruprecht\,130 VPD (see Appendix) exhibits a real concentration of stars and its CMD reveals unambiguous evolutionary sequences.

\subsubsection{NGC\,3519}

This OC is also named Ruprecht\,93 and it was analysed by \cite{Cheon:2010} by means of $UBVI$ CCD photometry. They were unable to confirm the existence of a real cluster from the spatial distribution of blue stars. The authors considered the absence of stars with $(U-B)_0<0.0$ (spectral types earlier than ~B9V; see their figures 1 and 2) in the central regions of NGC\,3519 as an indication that we are not facing a real OC. Here we propose a different interpretation. 

The bluer stars in NGC\,3519's CMD ($11.5\lesssim G\lesssim12.5\,$mag, $G_{\textrm{BP}}-G_{\textrm{RP}}<0.5\,$mag; see Appendix) present effective temperatures and masses around $\sim8500\,$K and $\sim2.5\,$M$_{\odot}$, respectively, as inferred from the PARSEC models. This parameters are more compatible with spectral type A. Consequently, no B stars are expected considering this cluster's age. Interestingly, in figure 2 of \cite{Cheon:2010} we can note that  the ZAMS relation reddened by $E(B-V)=0.25$ (which is reasonably close to our value) fits adequately a considerable number of stars with spectral types between A0$-$A7, compatible with NGC\,3519's turnoff (see Appendix). Despite this, the authors arbitrarily dismissed such stars as possible members. 

Consistently with \cite{Bica:2011}'s main conclusion, here we advocate that NGC\,3519 is a genuine OC, judging by the dispersion of data in its VPD and the evolutive sequences defined on its decontaminated CMD.

\begin{figure}
\begin{center}
 \includegraphics[width=8.0cm]{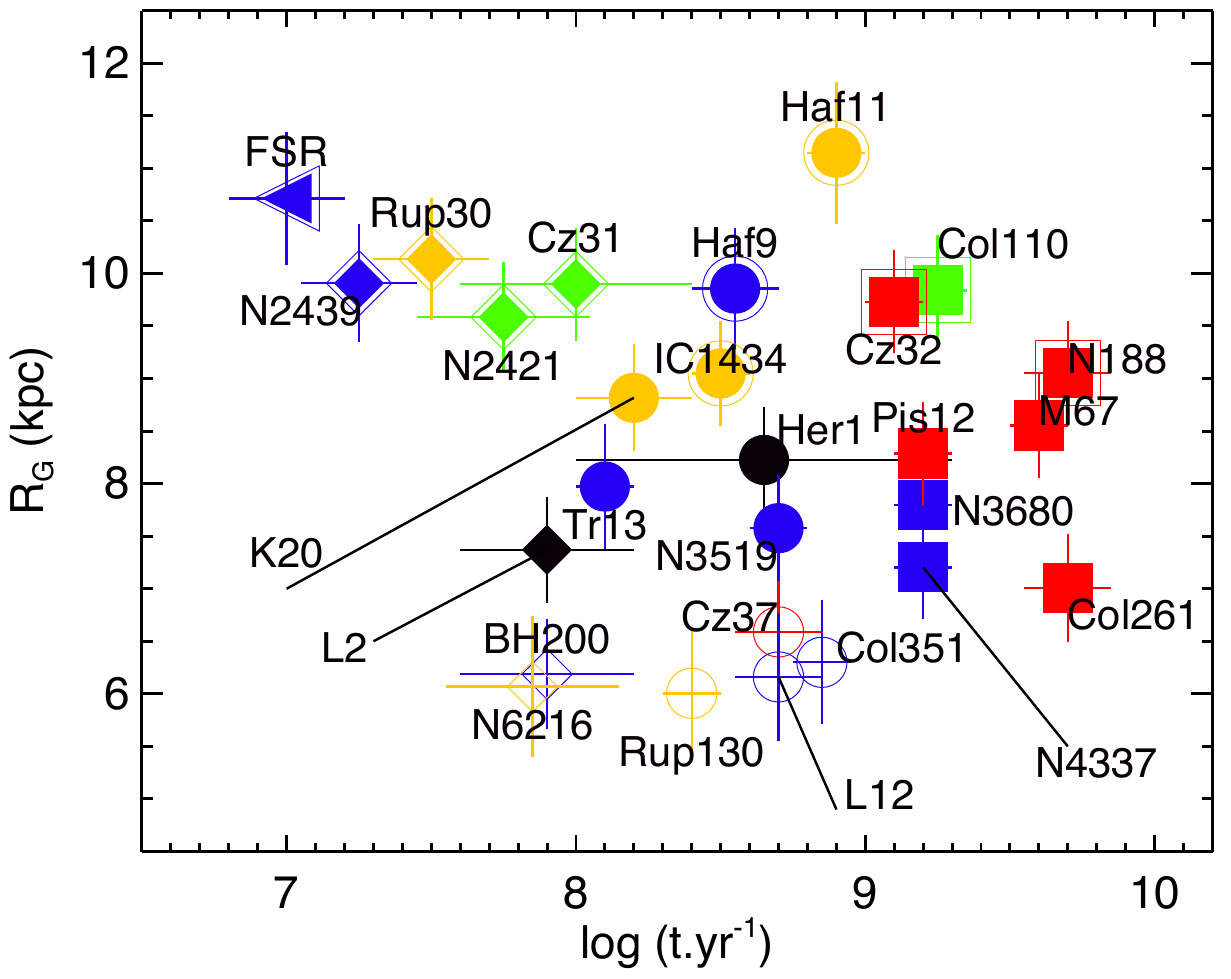}
 \caption{ Galactocentric distances ($R_G$) and ages for the investigated OCs. Symbols are given according to Table~\ref{symbols_convention} and acronyms for these clusters are shown. Lynga\,2 (L2) and Herschel\,1 (Her1), for which no $r_c$ and $r_{hm}$ could be derived (Table \ref{struct_params}), were plotted with black symbols. }
   \label{R_gal_versus_age_names}
\end{center}
\end{figure}

\section{Discussion}
\label{discussion}

OCs have their stellar content gradually depleted as they evolve dynamically. Internal relaxation transfers energy from the (dynamically) warmer central core to the cooler outer regions \citep{Portegies-Zwart:2010}, which causes low mass stars to be preferentially evaporated, thus becoming part of the general field population. Stars can also be ejected from the cluster after few encounters with hard binaries (e.g., \citeauthor{Heggie:1975}\,\,\citeyear{Heggie:1975}; \citeauthor{Hills:1975}\,\,\citeyear{Hills:1975}; \citeauthor{Hut:1983}\,\,\citeyear{Hut:1983}). External interactions can additionally cause the gradual dispersion of an OC stellar content, such as disc shocking (\citeauthor{Spitzer:1958}\,\,\citeyear{Spitzer:1958}; \citeauthor{Theuns:1991}\,\,\citeyear{Theuns:1991}) and collisions with giant molecular clouds (e.g., \citeauthor{Ostriker:1972}\,\,\citeyear{Ostriker:1972}). It is expected that the interplay between these destructive effects cause structural changes that can be probed from the relationships between $r_c$, $r_{hm}$ and $r_t$ (\citeauthor{Piatti:2017a}\,\,\citeyear{Piatti:2017a} and references therein).

In what follows, we show relations between structural and time-related parameters. We also present some general aspects of the investigated OCs.

\subsection{General characteristics of the investigated sample}

Fig.~\ref{R_gal_versus_age_names} exhibits the Galactocentric distances as a function of ages for our sample together with cluster acronyms. Symbols were assigned according to the following convention (see Table~\ref{symbols_convention}): open symbols for clusters with $R_G\leq7\,$kpc, filled symbols for those with $7<R_G (\textrm{kpc})\leq9$ and contoured symbols for $R_G>9\,$kpc. As explained in Table~\ref{symbols_convention}, our OCs were also categorized according to the age range (different symbols) and concentration parameter ($c$=log($r_t/r_c$)\,; different colours).

\begin{table}
\begin{minipage}{85mm}
  \caption{ symbol convention and colours used in Figs.~\ref{R_gal_versus_age_names} to \ref{plots_discussions_bloco3}. }
  \label{symbols_convention}
 \begin{tabular}{cccl}

\hline

  Age range                  &  $R_G\leq7\,$kpc    & $7<R_G\leq9\,$kpc    &  $R_G>9\,$kpc        \\ 
  log($t$.yr$^{-1}$)       &                                 &                                    &                                 \\
\hline 
       $\leq$\,7               &    $\lhd$                   &   $\blacktriangleleft$   & \hskip0.6cm\includegraphics[width=0.45cm]{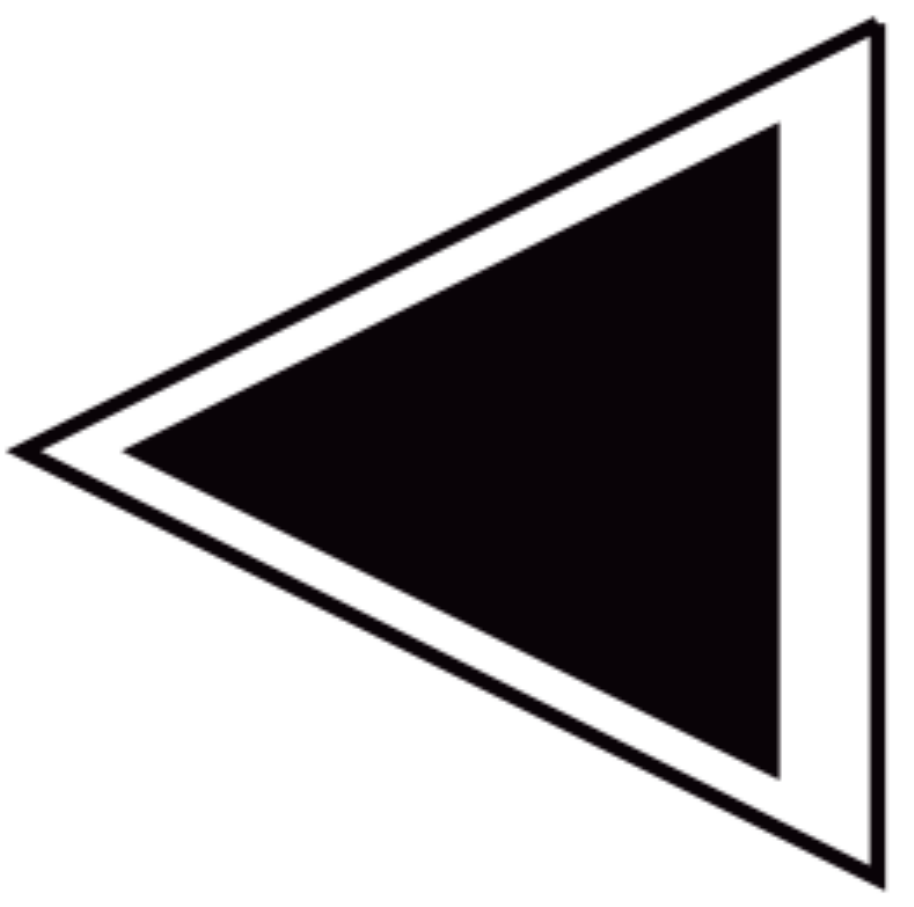} \\                     
\hline                                    
       $(7\,,\,8]$              &  $\lozenge$            & $\blacklozenge$         & \hskip0.6cm\includegraphics[width=0.45cm]{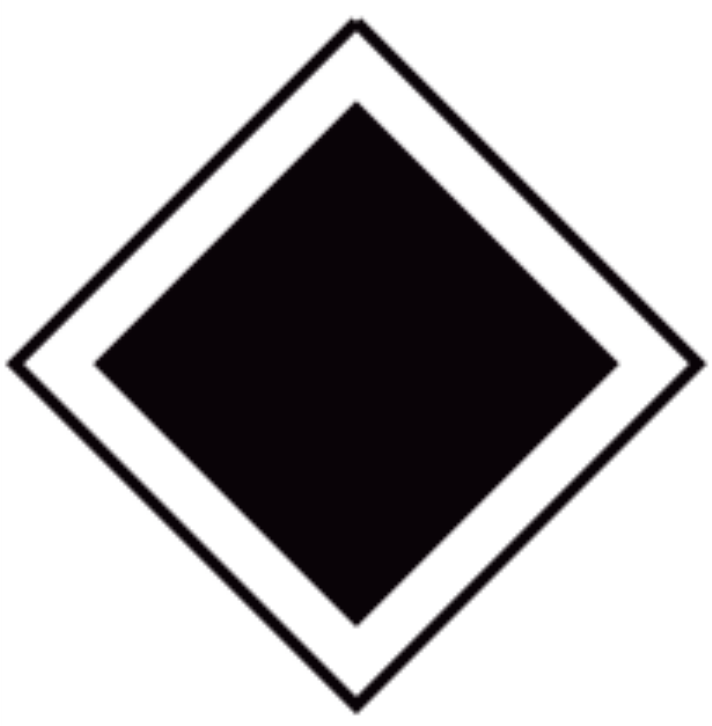}  \\
\hline                                           
       $(8\,,\,9]$              & \Large{$\circ$}        & \Large{$\bullet$}         & \hskip0.6cm\includegraphics[width=0.45cm]{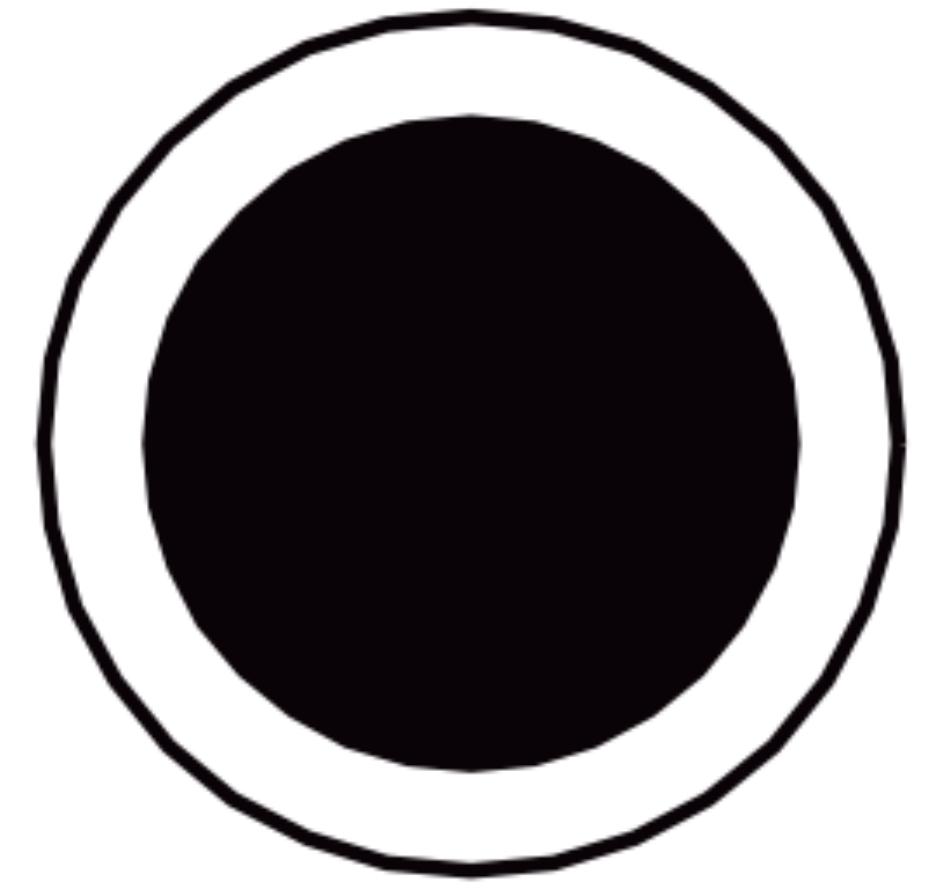} \\                 
\hline                                    
                                    
       $>\,$9                   & $\square$                & $\blacksquare$          & \hskip0.6cm\includegraphics[width=0.45cm]{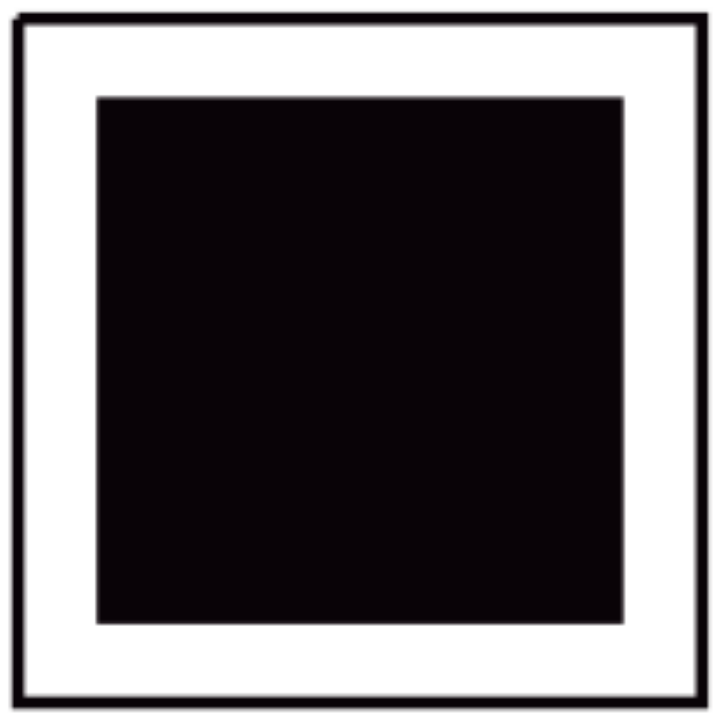} \\                  
                                   
\hline
\hline

Colour                      &  \multicolumn{3}{c}{$c$ (see Fig.~\ref{plots_discussions_bloco2}, panel a)}     \\

\hline

\textcolor{red}{Red}             &   \multicolumn{3}{c}{\textcolor{red}{$c\gtrsim0.70$}}                              \\
\textcolor{blue}{Blue}           &   \multicolumn{3}{c}{\textcolor{blue}{0.60$\lesssim c\lesssim0.70$}}      \\
\textcolor{orange}{Orange}  &   \multicolumn{3}{c}{\textcolor{orange}{$0.40\lesssim c\lesssim0.60$}}  \\
\textcolor{green}{Green}      &   \multicolumn{3}{c}{\textcolor{green}{$c\lesssim 0.40$}}  \\

\hline

\end{tabular}
\end{minipage}
\end{table}

\begin{figure*}
\begin{center}

\parbox[c]{1.0\textwidth}
  {
   \begin{center}
    \includegraphics[width=1.00\textwidth]{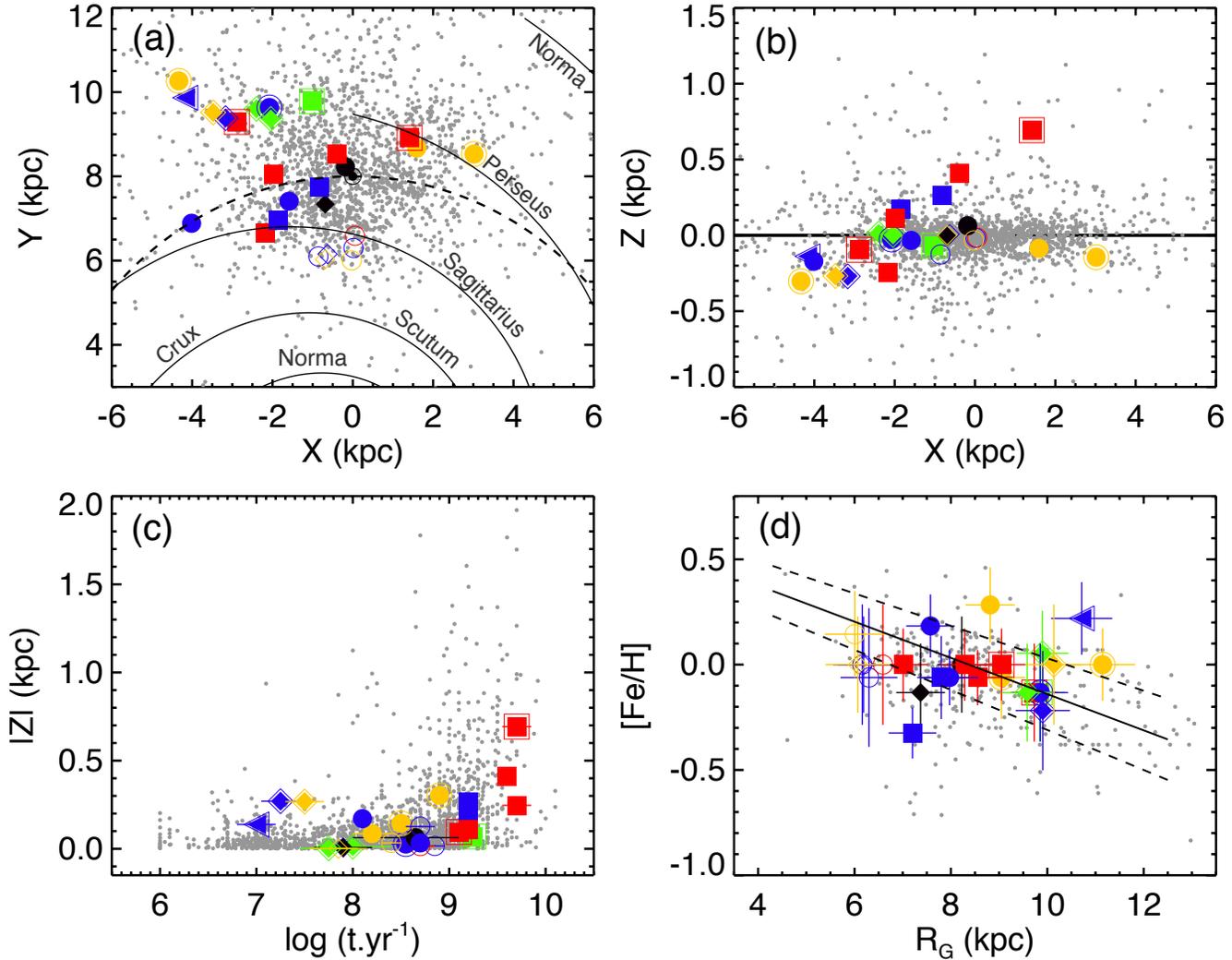}    
    \end{center}
    
  }
\caption{ Distribution of the investigated OCs along the Galactic plane (panel a) and perpendicular to the disc (panel b). The positions of the spiral arms were taken from \citeauthor{Vallee:2008}\,\,(\citeyear{Vallee:2008}). Panel (c) shows the Galactic Z coordinate as function of cluster age and panel (d) shows the dispersion of $[Fe/H]$ as function of $R_G$. For reference, the continuous line is the metallicity gradient fitted by Netopil et al. (2016); the uncertainties in this relation are represented as dashed lines. In all panels, small grey symbols represent OCs taken from the DAML02 catalogue.}

\label{plots_discussions_bloco1}
\end{center}
\end{figure*}

Fig.~\ref{plots_discussions_bloco1}, panel (a), shows that most of the investigated OCs are located between or close to the Sagittarius and Perseus arms. Their $R_G$ vary from $\sim6$ to 11\,kpc (Table \ref{struct_params}). The position of the Sun and the solar circle (dashed line) are represented. Panels (a) and (b) show the spatial distribution of our sample along the Galactic plane and perpendicular to the disc. Almost all investigated OCs are located within $\sim300\,$pc from the Galactic plane, which places them in the thin disc according to \cite{Chen:2001}, who derived a scale height of $z_h\sim330\,$pc for this Milky Way component; see also \cite{Gaia-Collaboration:2018a}, figures 2 and 11. Exceptions are M\,67 ($\vert Z\vert\sim400\,$pc) and NGC\,188 ($\vert Z\vert\sim700\,$pc). Panels (c) and (d) show how ages and metallicities distribute spatially. For comparison purposes, we also plotted clusters with available distances and ages from the DAML02 catalogue (small grey dots). 

The investigated clusters follow the general dispersion of data from the literature, with older clusters tending to be farther from the Galactic plane. $[Fe/H]$ values vary from $\sim-0.3$ to $\sim+0.3\,$dex, with most of the OCs in our sample presenting solar or close to solar metallic content. Again, they are consistent with the general dispersion of data in the $[Fe/H]\times\,R_{G}$ plot, as shown in panel (d). Considering uncertainties, most of them are located within the limits of the metallicity radial gradient ($d[Fe/H]/dR_{G}=-0.086\pm0.009\,$dex/kpc) as fitted by \citeauthor{Netopil:2016}\,\,(\citeyear{Netopil:2016}, continuous and dashed lines in panel d).

\subsection{Structural and time-related parameters}

\begin{figure*}
\begin{center}

\parbox[c]{1.0\textwidth}
  {
   \begin{center}
    \includegraphics[width=1.00\textwidth]{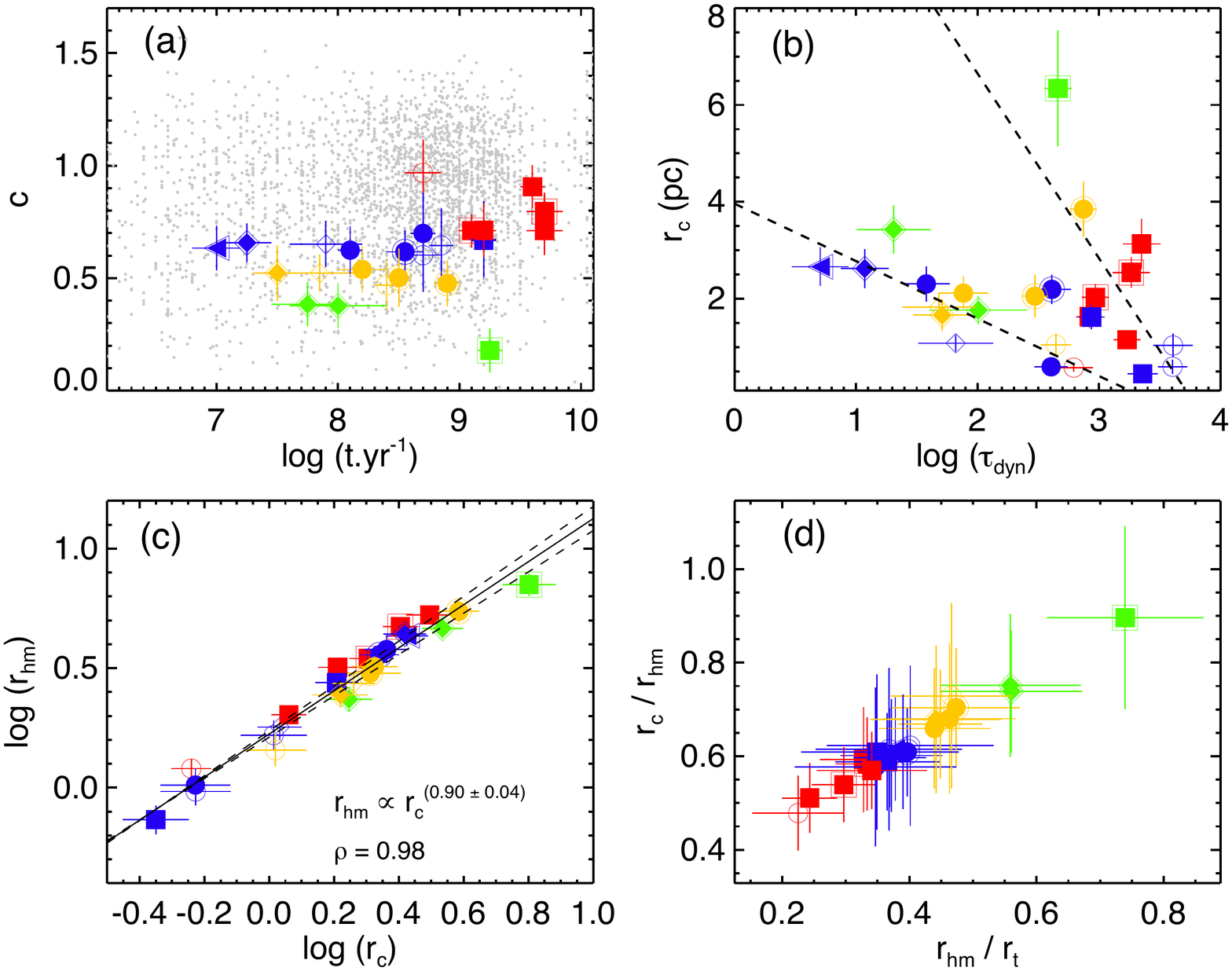}    
    \end{center}
    
  }
\caption{ Panel (a): Concentration parameter $c$ versus age plot. Symbol colours were given according to the $c$ intervals shown in Table~\ref{symbols_convention}. The grey dots represent clusters in the sample of Kharchenko et al. (2013). Panel (b): $r_c$ versus logarithmic dynamical ratio ($\tau_{\textrm{dyn}}$). The two dashed lines suggest different evolutionary paths and were plotted to guide the eye. Panel (c): $r_{hm}$ versus $r_c$ plot, both in log scale. The continous line is the linear fit between both radii and the dashed ones represent the fit uncertainties. Panel (d): $r_c/r_{hm}$ versus $r_{hm}/r_t$ plot. The same symbol convention was employed in all panels. }

\label{plots_discussions_bloco2}
\end{center}
\end{figure*}

\begin{figure*}
\begin{center}

\parbox[c]{1.0\textwidth}
  {
   \begin{center}
    \includegraphics[width=1.00\textwidth]{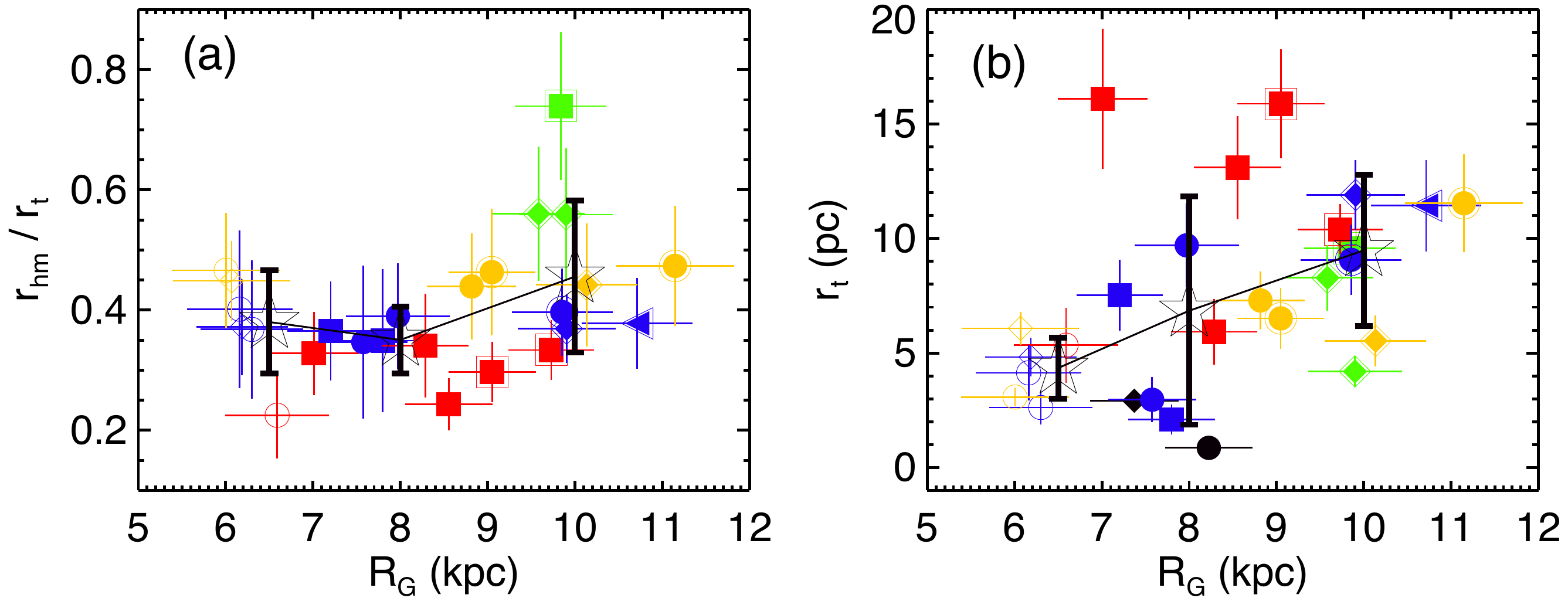}    
    \end{center}
    
  }
\caption{  Panel (a): $r_{hm}/r_t$ ratio versus Galactocentric distance ($R_G$). Panel (b): $r_t$ versus $R_G$ plot. The black symbols represent the OCs Lynga\,2 (filled diamond) and Herschel\,1 (filled circle), for which no $r_c$ and $r_{hm}$ could be obtained (see Table \ref{struct_params}). The black open stars show the mean and dispersion in $r_{hm}/r_t$ (panel a) and $r_t$ (panel b) for OCs in three $R_G$ bins (see text for details). }

\label{plots_discussions_bloco3}
\end{center}
\end{figure*}

Fig.~\ref{plots_discussions_bloco2}, panel (a), shows the concentration parameter as function of age for our OC sample and those in the literature (grey dots; from \citeauthor{Kharchenko:2013}\,\,\citeyear{Kharchenko:2013}, since DAML02 does not provide $r_t$ and $r_c$ values). The concentration parameter gives some evidence regarding the OC dynamical state, since internal relaxation causes higher mass stars to sink towards the cluster's centre and low-mass stars, preferentially, to evaporate. This conducts to a sucessively greater core-halo differentiation (e.g., \citeauthor{de-La-Fuente-Marcos:1997}\,\,\citeyear{de-La-Fuente-Marcos:1997}; \citeauthor{Portegies-Zwart:2001}\,\,\citeyear{Portegies-Zwart:2001}) and larger $c$ values. Compared to OCs in the literature, most of the OCs studied here are among those with smaller $c$ values for their ages.

The increasing degree of central concentration can be verified in Fig.~\ref{plots_discussions_bloco2}, panel (b), where $r_c$ is plotted as function of the dynamical ratio $\tau_{\textrm{dyn}}$ = age/$t_{\textrm{cr}}$. The crossing time ($t_{\textrm{cr}}$; see Table \ref{fundamental_params}) is a dynamical time-scale required for a typical star to cross the system. In this sense, the greater the $\tau_{\textrm{dyn}}$, the more dynamically evolved an OC. $t_{\textrm{cr}}$ can be defined as $t_{\textrm{cr}} = r_{hm}/\sigma_V$, where $\sigma_V$ is the intrinsic 3D velocity dispersion of the stars. The $\sigma_V$ values, converted to km\,s$^{-1}$ using the distances determined in Sect.~\ref{isoc_fitting}, were obtained from the dispersion in proper motions of member stars (Table \ref{fundamental_params}), assuming that the velocity components relative to each cluster centre are isotropically distributed. With this approximation, $\sigma_V=\sqrt{3/2}\,\sigma_{\mu}$, where $\sigma_{\mu}$ is the dispersion of the projected angular velocities ($\mu$), that is, $\mu=\sqrt{\mu_{\alpha}^2\,\textrm{cos}^2\,\delta\,+\,\mu_{\delta}^2}$. The uncertainties in the proper motions components were properly taken into account in the calculation of $\sigma_{\mu}$ by means of the procedure described in section\,4 of \cite{van-Altena:2013} and also in \cite{Sagar:1989}. 


In Fig.~\ref{plots_discussions_bloco2}, panel (b), we can note a general decrease in $r_c$ as a function of $\tau_{\textrm{dyn}}$, that is, clusters that have lived for many crossing times tend to present more compact central structures. Interestingly, the correlation between $r_c$ and $\tau_{\textrm{dyn}}$ seems to follow two different evolutionary paths in this plot, as suggested by the dashed lines (which were plotted here just to guide the eye). Particularly, for both trends suggested in panel (b) we can note a more distinct separation between clusters with $c\gtrsim0.70$ (red symbols) and those with $c\lesssim0.60$ (orange and green symbols): the more concentrated ones tend to present larger $\tau_{\textrm{dyn}}$. It is also noticeable from this panel that all clusters with log\,($\tau_{\textrm{dyn}}$)$>2.2$ are older than log($t\,$yr$^{-1})>8.3$.

The distribution of clusters with intermediate $c$ values (blue symbols) suggests that the OCs dynamical evolution is dominated by the internal two-body relaxation and regulated by the Galactic tidal field. In this context, the initial conditions at formation also play a role. Lynga\,12 and Collinder\,351 are part of the group of studied OCs with smaller Galactocentric distances and thus subject to a stronger Galactic gravitational pull, which may have accelerated their dynamical evolution. Both present log\,($\tau_{\textrm{dyn}})>3.6$ and may have suffered more severe mass loss. In turn, the old (log\,($t\,$yr$^{-1})>9$) clusters NGC\,4337 and NGC\,3680 present $c$ values compatible with those OCs with the highest concentrations (red symbols) and they are also among the group of more dynamically evolved ones (log\,($\tau_{\textrm{dyn}})\gtrsim3$). It is enlightening to verify that both present the same age and are placed at compatible R$_G$, but NGC\,3680 seems more evolved, presenting the smallest $r_c$ value in our sample. Consequently, differences in their present evolutionary stages may be linked to their properties at formation, like dissimilar original masses.

On the other hand, although [FSR2007]\,1325 and NGC\,2439 are young objects (log\,($t$\,yr$^{-1})<7.3$), their ages surpass the corresponding crossing times, therefore they present some degree of dynamical evolution.  Besides, their ages, Galactocentric distances and $\tau_{\textrm{dyn}}$ are similar within uncertainties, which makes them dynamically similar to each other. Consequently, they may possibly follow close evolutionary paths as they orbit the Galaxy. Other clusters plotted with blue symbols (namely, BH\,200, Haffner\,9, NGC\,3519 and Trumpler\,13) are in intermediate evolutionary stages. We also highlight the case of Collinder\,110, which is the less concentrated OC in our sample ($c=0.18$). Apparently, its relatively large Galactocentric distance ($R_G\sim10\,$kpc) allowed this OC to distribute its stellar content inside a less relaxed internal structure without being tidally disrupted, but on its way to dissolution.

Since the clusters are not age seggregated along the suggested sequences in panel (b) of Fig.~\ref{plots_discussions_bloco2} and neither distributed according to their Galactocentric distances, we speculate that both paths may be consequence of different initial formation conditions. Indeed, a spread of $r_c$ for a few Gyr old clusters in the Magellanic Clouds has been interpreted as the effect of black holes retained by the core as the cluster evolves \citep{Mackey:2008} or, alternatively, as due to a combination of different structure and physical properties at the moment of cluster birth and distinct stages of internal dynamical evolution \citep{Ferraro:2019}. In general, clusters with larger cores are dynamically less evolved than those with smaller cores. Analogous statements can be drawn for the half-mass radii, since $r_c$ and $r_{hm}$ are correlated as shown in panel (c) of Fig.~\ref{plots_discussions_bloco2}.


The level of internal relaxation can also be inferred from the $r_c/r_{hm}$ ratio. In fact, those with smaller $r_c/r_{hm}$ ratios present larger $c$ values. Following panel (b) of Fig.~\ref{plots_discussions_bloco2} and the discussions above, those more centrally concentrated (i.e., smaller $r_c/r_{hm}$) tend to be in more advanced evolutionary stages. In turn, the $r_{hm}/r_t$ ratio can be used as a probe of the tidal influence of the Galaxy on the cluster's dynamical evolution \citep{Baumgardt:2010}. Clusters for which their half-mass content fills a larger fraction of the tidal volume are subject to a larger mass loss due to tidal effects. Panel (d) of Fig.~\ref{plots_discussions_bloco2} shows that $r_c/r_{hm}$ and $r_{hm}/r_t$ ratios present a positive correlation, that is, those more compact are in fact less subject to disruption by the Galactic tidal field. It is noticeable that the data along the suggested trend are not seggregated according to OCs ages or Galactocentric distances.

The set of results presented in Fig.~\ref{plots_discussions_bloco2} suggests an overall disruption scenario in which OCs tend to become more compact structures as they evolve dynamically. This means that, in the far future, initially massive OCs will leave behind only a compact core, with most of their low-mass stellar content dispersed in the general Galactic field (\citeauthor{de-La-Fuente-Marcos:1996}\,\,\citeyear{de-La-Fuente-Marcos:1996}; \citeauthor{Bonatto:2004a}\,\,\citeyear{Bonatto:2004a}).

\subsection{Differential tidal effects on OCs' structure}

Panel (a) of Fig.~\ref{plots_discussions_bloco3} allows to evaluate if the $r_{hm}/r_t$ ratio is differentially affected by the Galactic tidal field. The positions of the black open stars in the plot and their error bars represent, respectively, the mean and standard deviation of the $r_{hm}/r_t$ ratio for OCs in three $R_G$ bins: $R_G\leq7\,$kpc, $7<R_G (\textrm{kpc})\leq9$ and $R_G>9\,$kpc. There is not a clear trend with $R_G$, but the dispersion in $r_{hm}/r_t$ for OCs in the range $R_G>9\,$kpc is at least 50\% larger compared to the other $R_G$ bins. The smaller dispersion in $r_{hm}/r_t$ for smaller $R_G$ may be due to the fact that these OCs are subject to a stronger Galactic tidal field. In these cases, smaller $r_{hm}/r_t$ values favor their survival against tidal disruption.

Changes in the Galactic gravitational potential may have imprinted differential tidal effects in the OCs outermost structures. Panel (b) of Fig.~\ref{plots_discussions_bloco3} shows, on average, an increase in $r_t$ as function of the Galactocentric distance, considering the same three $R_G$ bins of panel (a). The dispersions in $r_t$ are also considerably larger for those located at $R_G\gtrsim7\,$kpc compared to those closer to the Galactic centre (open coloured symbols). OCs located at $R_G\lesssim7\,$kpc are subject to a stronger external gravitational field, which may explain their more compact external structures.

\section{Summary and concluding remarks}
\label{conclusions}

In the present paper we revisited the cases of 14 objects previously classified as low contrast or faint Galactic OCs. In general, these OCs are projected against a dense stellar background, due to their low Galactic latitudes (typically $\vert b\vert\lesssim5^{\circ}$), and/or they present depleted evolutionary sequences, due to mass loss as a consequence of their dynamical evolution. Other 2 OCs (namely, NGC\,2421 and [FSR2007]\,1325) were included in the main sample, since they are projected in the same fields of Czernik\,31 and Czernik\,32. In fact, our results suggest that Czernik\,31 and NGC\,2421 form a physical binary, as inferred from their compatible distances, colour excesses and RDPs. The dynamics of this probable pair could be better investigated with the determination of radial velocities. In turn, the apparent proximity between Czernik\,32 and [FSR2007]\,1325 seems just a projection effect, since these two clusters present distances and colour excesses that are incompatible with the hypothesis of a pair of interacting OCs. For comparison purposes, we also investigated a complementary sample of 11 OCs, which are more readily distinguishable from the general Galactic disc population.

For each cluster, our analysis procedure was divided in three basic steps: $(i)$ a pre-analysis procedure, during which the applying of dedicated magnitude filters on the cluster CMD and the inspection of its VPD allowed to check preliminarly its physical existence; at this stage, a proper motions filter was also applied in order to alleviate the contamination by field stars; $(ii)$ determination of structural parameters from the cluster RDP by minimisation of $\chi^2$ while fitting \cite{King:1962}'s and \cite{Plummer:1911}'s profiles; $(iii)$ selection of member stars: at this stage, the magnitude filter was dismissed and we searched for stars with high membership likelihoods in the region $r\leq r_t$; these stars define clear evolutionary sequences in decontaminated $G\times (G_{\textrm{BP}}-G_{\textrm{RP}})$ CMDs. 

All investigated objects had their physical nature confirmed as genuine OCs, contrarily to studies that pointed out the presence of some asterisms, which were the cases of Ruprecht\,30 and NGC\,3519. In the case of Ruprecht\,130, no firm conclusions could be drawn by previous studies regarding its physical nature. Our results point out a real physical system. Mean astrometric parameters derived here are in close agreement with recent studies based on GAIA DR2. Despite this, the fundamental astrophysical parameters $E(B-V)$, $(m-M)_0$ and log\,$(t)$ present important discrepancies with the literature, more probably due to different procedures of member stars selection. Since our method is based on high-precision astrometry and photometry, we could obtain deeper CMDs and more refined lists of members.      


Our investigated OCs are located between or close to the Sagittarius and Perseus spiral arms ($R_G$\,$\sim6-$11\,kpc), spanning a wide age range (log\,$t\sim$7.0$-$9.7) and evolutionary stages. Most of them present solar metallic content. The joint analysis of our complete sample suggests a general scenario in which, as clusters evolve dynamically and gradually lose stars (preferentially lower mass ones) by evaporation, their overall structure become more compact. This can be inferred from the negative correlation between $r_c$ and the dynamical ratio $(\tau_{\textrm{dyn}}$), that is, more centrally concentrated clusters tend to be more dynamically evolved. The trends exhibited in the $r_c$ versus $\tau_{\textrm{dyn}}$ plot suggest two distinct evolutionary sequences, which may be consequence of different initial conditions at formation. From the positive correlation between $r_c/r_{hm}$ and $r_{hm}/r_t$ ratios, we can notice that in fact more compact OCs are less tidally filled and thus less subject to a larger mass loss due to tidal effects. In this scenario, for a given set of initial formation conditions, the OC dynamical evolution is ruled by internal two-body relaxation, regulated by the external tidal field.

The degree of tidal filling, as measured by the $r_{hm}/r_t$ ratio, does not exhibit a clear correlation with $R_G$, but this ratio presents smaller dispersion for OCs closer to the Galactic centre. Since these structures are subject to a stronger Galactic gravitational pull, their smaller $r_{hm}/r_t$ values favour their survival against tidal disruption. The OCs' external structure seems differentially affected by changes in the external gravitational potential: for those located at smaller $R_G$, their tidal $r_t$ are smaller and with less dispersion compared to those at larger $R_G$.

The availability and spatial coverage of data in the GAIA DR2 catalogue, combined with analyses methods such as those proposed in the present paper, allows uniform and refined characterizations of an increasingly large number of clusters. This is an essential step towards a deep understanding of the Galactic structure and its evolution, as well as to a better comprehension of the clusters disruption process.

\section{Acknowledgments}

We thank the anonymous referee for helpful suggestions. This research has made use of the VizieR catalogue access tool, CDS, Strasbourg, France. This work has made use of data from the European Space Agency (ESA) mission Gaia (https://www.cosmos.esa.int/gaia), processed by the Gaia Data Processing and Analysis Consortium (DPAC, https://www.cosmos.esa.int/web/gaia/dpac/consortium). Funding for the DPAC has been provided by national institutions, in particular the institutions participating in the Gaia Multilateral Agreement. This research has made use of \textit{Aladin sky atlas} developed at CDS, Strasbourg Observatory, France. The authors thank the Brazilian financial agencies CNPq and FAPEMIG. This study was financed in part by the Coordena\c{c}\~ao de Aperfei\c{c}oamento de Pessoal de N\'ivel Superior $-$ Brazil (CAPES) $-$ Finance Code 001.


\bibliographystyle{mn2e}
\bibliography{referencias}




\bsp

\label{lastpage}

\end{document}